\documentclass[journal]{IEEEtran}
\usepackage{graphicx}
\usepackage{subfigure}
\usepackage{booktabs} 
\usepackage{amsmath}
\usepackage{textcomp}
\usepackage{listings}
\usepackage{xcolor}
\usepackage{threeparttable}
\usepackage{multirow}
\usepackage{array,booktabs}
\graphicspath{ {images/}}
\usepackage[hidelinks]{hyperref}
\usepackage{url}
\usepackage{cite}
\usepackage{algorithm}
\usepackage{algpseudocode}

\definecolor{codeblack}{rgb}{0, 0, 0}
\definecolor{codegreen}{rgb}{0.2, 0.5, 0.2}

\lstdefinestyle{mystyle}{
    commentstyle=\itshape\color{codegreen},
    keywordstyle=\color{magenta},
    basicstyle=\ttfamily\footnotesize\color{codeblack},
    breakatwhitespace=false,         
    breaklines=true,                 
    captionpos=b,                    
    keepspaces=true,                 
    numbersep=5pt,                  
}

\ifCLASSINFOpdf
\else
\fi
\begin{document}
%
\title{EasyFL: A Low-code Federated Learning Platform For Dummies}
%
%
%


\author{Weiming~Zhuang,~\IEEEmembership{Student Member,~IEEE,}
        Xin~Gan, 
        Yonggang~Wen,~\IEEEmembership{Fellow,~IEEE,}
        and~Shuai~Zhang
\thanks{W. Zhuang is with S-Lab, Nanyang Technological University, Singapore 639798 (email: weiming001@e.ntu.edu.sg).}
\thanks{X. Gan, and Y. Wen are with Nanyang Technological University, Singapore 639798 (email: ganx0005@e.ntu.edu.sg; ygwen@ntu.edu.sg).}
\thanks{W. Zhuang and S. Zhang are with SenseTime Research, China (e-mail: zhuangweiming@senetime.com; zhangshuai@sensetime.com).}
\thanks{Copyright (c) 2022 IEEE. Personal use of this material is permitted. However, permission to use this material for any other purposes must be obtained from the IEEE by sending a request to pubs-permissions@ieee.org.}}

\maketitle

\begin{abstract}

Academia and industry have developed several platforms to support the popular privacy-preserving distributed learning method -- Federated Learning (FL). However, these platforms are complex to use and require a deep understanding of FL, which imposes high barriers to entry for beginners, limits the productivity of researchers, and compromises deployment efficiency. In this paper, we propose the first low-code FL platform, \textit{EasyFL}, to enable users with various levels of expertise to experiment and prototype FL applications with little coding. We achieve this goal while ensuring great flexibility and extensibility for customization by unifying simple API design, modular design, and granular training flow abstraction. With only a few lines of code, EasyFL empowers them with many out-of-the-box functionalities to accelerate experimentation and deployment. These practical functionalities are heterogeneity simulation, comprehensive tracking, distributed training optimization, and seamless deployment. They are proposed based on challenges identified in the proposed FL life cycle. Compared with other platforms, EasyFL not only requires just three lines of code (at least 10x lesser) to build a vanilla FL application but also incurs lower training overhead. Besides, our evaluations demonstrate that EasyFL expedites distributed training by 1.5x. It also improves the efficiency of deployment. We believe that EasyFL will increase the productivity of researchers and democratize FL to wider audiences.
\end{abstract}



%
\IEEEpeerreviewmaketitle

\begin{table*}[t]
\caption{Compare our proposed EasyFL with existing FL libraries, platforms, and frameworks. EasyFL requires little coding and provides more out-of-the-box functionalities to improve the productivity of researchers. $\bigcirc$ means limited support.}
\label{tab:work-comparison}
\begin{center}
\begin{tabular}{lcccccc}
\hline
 & LEAF \cite{leaf} & PySyft \cite{pysyft} & PaddleFL \cite{ma2019paddlepaddle} & TFF \cite{tff} & FATE \cite{fate} & EasyFL (Ours) \\
\hline
Lines of Code (Vanilla FL App) & $\sim$400 & $\sim$190 & $\sim$190 & $\sim$30 & $\sim$100 & 3 \\
Heterogeneity Simulation & $\bigcirc$ & $\times$ & $\bigcirc$ & $\bigcirc$ & $\bigcirc$ & $\surd$ \\
Training Flow Abstraction  & $\times$ & $\times$ & $\times$ & $\times$ & $\times$ & $\surd$ \\
Distributed Training Optimization & $\times$ & $\times$ & $\times$ & $\times$ & $\times$ & $\surd$ \\
Tracking & $\times$ & $\times$ & $\times$ & $\bigcirc$ & $\surd$ & $\surd$ \\
Deployability & $\times$ & $\times$ & $\surd$ & $\bigcirc$ & $\surd$ & $\surd$ \\
\hline
\end{tabular}
\end{center}
\end{table*}

\section{Introduction}

\IEEEPARstart{F}{ederated} learning (FL) is attracting considerable attention in recent years. It is a new distributed training technique that provides in-situ model training on decentralized edges. FL has empowered a wide range of applications with privacy-preserving mechanism, including healthcare applications \cite{sheller2018brain-tumor2, li2019brain-tumor1, chen2020fedhealth}, consumer products \cite{hard2018gboard, leroy2019kwakeword}, recommendation systems \cite{niu2020recommendation-alibaba, muhammad2020fedfast}, and video surveillance \cite{zhuang2020fedreid}. Companies and institutions are exploring these applications mostly in experimental environments \cite{li2019brain-tumor1, muhammad2020fedfast}. Some of them are building prototypes and then deploying FL applications \cite{yang2018gboard, rieke2020nvidia-fl-health}.

However, AI engineering \cite{gartner} of FL requires tremendous resources and efforts, regardless of levels of expertise. Beginners who are interested in FL face high barriers to entry due to the complex setup and non-trivial concepts. Researchers who lack practical experience in FL would need two to three weeks to build a vanilla FL application from scratch. More advanced features like comprehensive tracking and application-specific optimizations are even more time-consuming, hindering their productivity. Experienced researchers who studied FL would face challenges when building prototypes and deploying FL applications because they may not be familiar with software engineering and infrastructure. Communication and collaboration with software engineers who do not know FL would be a hassle.

Existing FL platforms and frameworks from academia and industry are complex to use and require a deep understanding of FL. Most existing FL systems need at least 100 lines of code to implement a vanilla FL application, as shown in Table \ref{tab:work-comparison} (links of source codes are provided in Appendix \ref{sec:appx-application}). LEAF \cite{leaf} is the first FL benchmark, but users need to replicate and learn the whole library to start experimentation. Although TensorFlow Federated (TFF) \cite{tff} is a research-oriented platform with two layers of APIs, it imposes prerequisite on knowledge in Tensorflow \cite{abadi2016tensorflow} and Keras \cite{chollet2015keras}. PySyft \cite{pysyft} focuses on secure and private deep learning, which is not straightforward for users interested in FL. All these platforms lack the deployability to build FL prototypes and run FL in production. Industrial FL frameworks such as Federated AI Technology Enabler (FATE) \cite{fate} and PaddleFL \cite{ma2019paddlepaddle} are deployable with containerization, but they are unfriendly to beginners and researchers due to complex environment setup and heavy system design.

In this paper, we propose a new paradigm of FL system, a \textit{low-code} FL platform termed \textit{EasyFL}, to enable users with various levels of expertise to experiment and prototype FL applications with little coding. Using our platform, beginners can start experiencing FL with only three lines of code, even without prior knowledge of FL. For researchers who are addressing the advanced and open problems in FL \cite{kairouz2019fl-advances-open}, we provide them with sufficient flexibility to develop new algorithms with minimal coding by reusing the majority of FL architecture. EasyFL empowers experienced researchers to easily prototype and seamlessly deploy FL applications without further engineering.



Although users write less code, we analyze the challenges in the FL life cycle and empower them with more out-of-box functionalities. We summarize and compare these functionalities with other platforms in Table \ref{tab:work-comparison}. Specifically, to facilitate ease of experimentation and development, EasyFL supports heterogeneity simulation, training flow abstraction, and comprehensive tracking: 1) The out-of-the-box heterogeneity simulation, only partially supported by other platforms, enables researchers to easily start exploring the most important challenges in FL --- statistical heterogeneity and system heterogeneity \cite{Li2020FedChallenges}; 2) Instead of using logs to collect results like most other platforms, we design a hierarchical tracking system to organize training metrics for easier result analysis; 3) EasyFL is the first platform that abstracts FL training flow to granular stages for minimum coding to develop new algorithms. Moreover, we exclusively support distributed training optimization to accelerate and scale FL training. Last but not least, EasyFL provides seamless and scalable deployment of centralized topology-based FL systems with containerization and service discovery, whereas existing platforms either focus on the deployment of multi-party FL (FATE \cite{fate} and PaddleFL \cite{ma2019paddlepaddle}) or have incomplete deployment features (TFF \cite{tff}).


Our evaluation demonstrates that EasyFL effectively reduces the lines code for developing FL applications and accelerates experimentation and deployment. We implement three FL applications with 4.5x to 9.5x fewer codes than the original implementations. Despite that we provide many out-of-the-box functionalities through abstractions, EasyFL has a lower training overhead compared with other popular FL libraries. EasyFL further accelerates distributed training by 1.5x under limited hardware resources. It also allows sub-linear time to deployment and maintains performance in production. We believe that EasyFL will lower the barriers to entry for beginners, increase the productivity of researchers, and bridge the gap between researchers and engineers.


In summary, we make the following contributions:
\begin{itemize}
    
    \item We define the FL life cycle and propose the system requirements by analyzing challenges in the life cycle.
    
    \item We design and build the first \textit{low-code} FL system, EasyFL, to democratize FL to wider audiences. Complex system implementations are shielded for users, regardless of expertise levels. We unify simple API design and granular training flow abstractions to achieve the low-code capability. Besides, we further facilitate ease of experimentation and development with out-of-the-box heterogeneity simulation and comprehensive tracking.
        
    \item  We provide sufficient flexibility and extensibility through abstraction and plugin architecture for experienced researchers to easily develop new algorithms and applications to address open problems in FL.
    
    \item We propose distributed training optimization to accelerate and scale FL experimentation.
    
    \item We support seamless and scalable deployment for fast prototyping and training in production.
    

\end{itemize}

The rest of the paper is organized as follows. In Section \ref{sec:related-work}, we introduce the background of FL and review related work. We define and analyze the FL life cycle to illustrate the system requirements in Section \ref{sec:workflow}.
Section \ref{sec:architecture} presents the high-level overview of system design and architecture of EasyFL. In Section \ref{sec:easy-development}, we introduce the modules to facilitate ease of experimentation and development. We propose distributed training optimization in Section \ref{sec:distributed-training-optim} and present seamless and scalable deployment in Section \ref{sec:deployment}. In Section \ref{sec:experiments}, we provide the evaluation of EasyFL. Section \ref{sec:conclusion} summarizes this paper and provides future directions.


\begin{figure*}[t]
    \centering
    \includegraphics[width=0.79\textwidth]{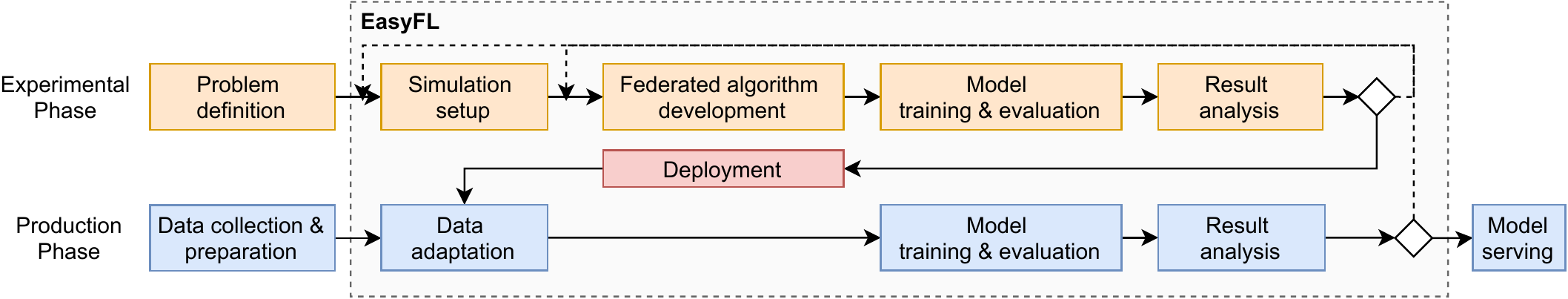}
    \caption{Illustration of key activities in two phases of the federated learning life cycle. In the experimental phase, researchers conduct experiments and develop new algorithms to solve problems in simulated environments. After deploying to the production, researchers use those algorithms to train models with real-world data in the actual environment.}
    \label{fig:fl-workflow}
\end{figure*}

\section{Background and Related Work}
\label{sec:related-work}

Federated learning (FL) is a distributed learning paradigm where multiple clients train machine learning (ML) models with the coordination of a central server \cite{kairouz2019fl-advances-open}. Federated Averaging (FedAvg) \cite{fedavg} is a standard FL algorithm. It completes the training with the following three steps: (1) the server selects a fraction of clients and sends them a global model $w$; (2) each client $k$ trains the model using their datasets for $E$ local epochs and uploads updates $w_k$. (3) the server aggregates these updates to obtain a new global model. EasyFL implements FedAvg as the default algorithm.

\textbf{Federated Learning Platforms} Institutions and companies have developed several experimentation-oriented and industrial-level FL platforms and frameworks \cite{leaf, pysyft, ma2019paddlepaddle, tff, fate}. Although experimentation-oriented libraries like LEAF \cite{leaf} provide sample implementations and simulate statistical heterogeneity with federated datasets, they are lack of deployability. TFF \cite{tff} is a deployable research-oriented platform, but it does not optimize distributed training. Although industrial-level frameworks like FATE \cite{fate} supports deployment with containers, they are not user-friendly due to high barriers of entry caused by heavy system design. Our proposed low-code platform, EasyFL, is user-friendly and supports efficient experimentation and seamless deployment.

\textbf{Heterogeneity in FL} Statistical and system heterogeneity are the key challenges of FL \cite{Li2020FedChallenges}. Statistical heterogeneity is sourced from non-independent and identical distributed (non-IID) and unbalanced data. Many studies propose novel approaches to address it using federated datasets \cite{zhao2018non-iid, zhuang2020fedreid, fedprox}. System heterogeneity is caused by varieties of hardware and networking conditions of clients. Exiting system heterogeneity simulations are either hardware-based \cite{tifl} that are complex and resource-demanding or software-based \cite{flash, fedprox} that are not generic enough. EasyFL simulates statistical heterogeneity with the three most commonly used datasets (Table \ref{tab:datasets-models}) and simulates system heterogeneity in a lightweight manner.

\textbf{Distributed ML} Distributed ML is a common practice to train a large-scale model in data centers, using model parallelism and data parallelism \cite{large-scale-dnn-NIPS2012}. FL differs from distributed ML because it faces unique challenges of expensive communication, statistical and system heterogeneity, and privacy concerns \cite{Li2020FedChallenges}. In experimentation, FL training can allocate each selected client to one GPU to boost the training speed, which is similar to data parallelism. EasyFL accelerates training by dynamically allocating multiple clients to a GPU under resource constraints and heterogeneity simulation.

%


\section{Federated Learning Life Cycle}
\label{sec:workflow}

In this section, we propose the FL life cycle based on in-depth studies of existing research and industrial practices. In addition, we present the system requirements by analyzing the challenges of activities in the life cycle.


\subsection{Life Cycle}
\label{sec:fl-workflow}

The life cycle of FL comprises two phases: the experimental phase and the production phase. Fig. \ref{fig:fl-workflow} presents key activities in both phases. Researchers develop novel algorithms to address specific problems in simulation-based settings in the experimental phase. After deployment, they evaluate the algorithms under real-world scenarios in the production phase. We propose this FL life cycle based on in-depth studies of research and industrial practices of FL \cite{kairouz2019fl-advances-open, fate} and traditional machine learning engineering for production (MLOps) \cite{ml-debt, burkov2020mle-book,zhang2020mlmodelci,karlavs2020mlops-kdd}. However, different from MLOps that deploys models to perform inference for new requests, FL trains new models with real-world decentralized data in the production phase.

In the experimentation phase, researchers solve problems by iterating these steps: (1) Identify the problems; (2) Setup the simulation environments; (3) Adapt existing algorithms and develop new algorithms; (4) Conduct experiments with different combinations of hyperparameters; (5) Analyze results. 
  
The key step to transit from the experimental phase to the production phase is deployment. 
The production phase contains the following steps: (1) Collect and transform data in clients; (2) Adapt datasets to FL systems; (3) Train and evaluate the algorithm developed in the experimentation phase; (4) Analyze results; (5) Publish the model for serving. 


In both phases, if the training results are not promising, researchers usually consider two ways to improve performance: (1) Starting from the simulation setup again to ensure that simulated scenarios match real-world scenarios; (2) Iterating algorithms if the simulation setup meets the requirement.

\subsection{System Requirements}
\label{sec:design-requirement}

Based on the FL life cycle, we present the system requirements of a low-code FL platform by analyzing the challenges of the critical activities in the life cycle. 




\textbf{Ease of Experimentation and Development} Fundamentally, an FL platform must allow users with different levels of expertise to easily run FL experiments, develop new algorithms, and implement new applications. To achieve this objective, the platform must resolve the following challenges. Firstly, statistical and system heterogeneity are the most common challenges: simulation of statistical heterogeneity requires methods to partition datasets to become non-IID and unbalanced; simulation of system heterogeneity (tightly related to hardware) currently is complex and resource-demanding. 

Secondly, most of the recently innovated federated algorithms, such as FedProx \cite{fedprox} and FedReID \cite{zhuang2020fedreid}, change only a few aspects of the existing algorithms (more examples are provided in Appendix \ref{sec:appx-publications}). Replicating the whole codebase for FL training would be a waste of time and effort. 

Thirdly, FL metrics collection is not trivial because it needs to support: (1) diverse training methods; (2) metrics in three levels of hierarchy: a training task comprises metrics of rounds where a round contains metrics of clients. Existing logging methods like TensorBoard \cite{abadi2016tensorflow} cannot fulfill these requirements.

To address these challenges, EasyFL provides out-of-the-box heterogeneity simulation to simulate FL challenges with the least hassles, abstracts the training flow to enable minimum coding for new algorithms and applications, and supports comprehensive tracking for in-depth analysis (Section \ref{sec:easy-development}).

\textbf{Diverse Training Methods} An FL platform must support three training methods: standalone, distributed, and remote training. In the experimental phase, \textit{standalone training} simulates FL on a hardware device (CPU or GPU), which is the simplest way to start FL experiments. \textit{Distributed training} increases training speed by training on multiple hardware devices, using communication backend provided by frameworks like TensorFlow \cite{abadi2016tensorflow}. Existing distributed training allocates one client to a GPU \cite{fedprox}. Further optimization is needed because 1) available hardware resources are normally less than the number of simulated clients (tens or hundreds); 2) simulations of unbalanced data or system heterogeneity lead to stragglers, resulting in resource underutilization when fast clients wait for the stragglers. 

Moreover, in the production phase, \textit{remote training} is needed to support remote message transmission as the server and clients are deployed to different locations. An FL platform must integrate remote communication into the training flow, without affecting standalone and distributed training. 

EasyFL supports these three training methods by decoupling the communication channels and the training flow. It provides standalone training by default, optimizes distributed training (Section \ref{sec:distributed-training-optim}), and supports remote training by transmitting messages through remote communication (Section \ref{sec:deployment}).

\textbf{Seamless and Scalable Deployment} Deployment, the last step to run FL in production, is also challenging because of diverse computing environments and the potentially large number of clients. Firstly, the production environment of FL clients could have large variances on the hardware and operating systems \cite{Li2020FedChallenges}. Manually adapting and deploying is time-consuming and unscalable. Moreover, new FL algorithm development usually needs multiple iterations of deployment because the algorithm verified in the experimental environment may not work perfectly well in the production environment. Secondly, to scale up the number of clients in training in production, the server must be aware of the addresses of clients. Using static configuration is a straightforward approach to manually record addresses of clients on the server. However, this method is unstable and unscalable when the number of clients is large because it is common that existing clients would drop out and new clients would join \cite{Bonawitz2019FL-sys-scale}. 

To tackle these challenges, EasyFL supports seamless and scalable deployment with containerization and service discovery (Section \ref{sec:deployment}).

\section{System Design and Architecture}
\label{sec:architecture}

\begin{figure}[t]
    \centering
    \includegraphics[width=0.47\textwidth]{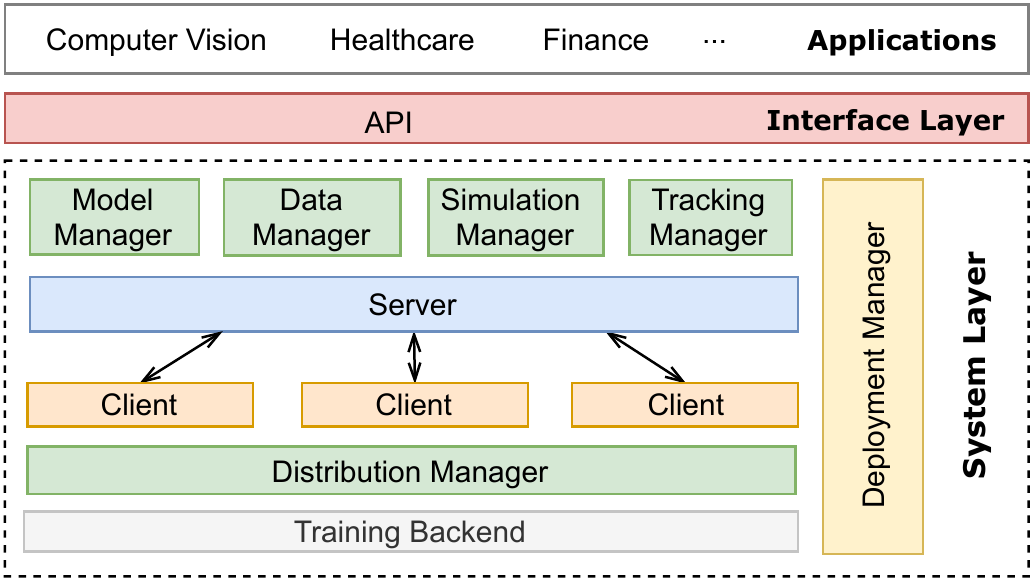}
    \caption{EasyFL's architecture consists of two parts: an interface player providing simple APIs for applications and a modularized system layer providing flexibility, extensibility, and out-of-the-box functionalities.}
    \label{fig:architecture}
\end{figure}

In this section, we first provide a high-level overview of the system design and architecture of EasyFL (Fig. \ref{fig:architecture}). Then, we present simple designs of APIs (Table \ref{tab:interfaces}).




\subsection{Overview}

EasyFL's architecture comprises of an \textit{interface layer} and a modularized \textit{system layer}. The interface layer provides simple APIs for high-level applications and the system layer has complex implementations to accelerate training and shorten deployment time with out-of-the-box functionalities.


\textbf{Interface Layer} The interface layer, providing a common interface across FL applications, is one of the key designs that enable low-code capability. It contains APIs that are designed to encapsulate complex system implementations from users. These APIs decouple application-specific models, datasets, and algorithms such that EasyFL is generic to support a wide range of applications like computer vision and healthcare. 

\textbf{System Layer} The system layer supports and manages the FL life cycle. It consists of eight modules to support FL training pipeline and life cycle:
1) The \textit{simulation manager} initializes the experimental environment with heterogeneous simulations. 2) The \textit{data manager} loads training and testing datasets, and the \textit{model manager} loads the model. 3) A \textit{server} and the \textit{clients} start training and testing with FL algorithms such as FedAvg \cite{fedavg}. 4) The \textit{distribution manager} optimizes the training speed of distributed training. 5) The \textit{tracking manager} collects the evaluation metrics and provides methods to query training results. 6) The \textit{deployment manager} seamlessly deploys FL and scales FL applications in production.

The system layer embraces coarse-grained plugin architecture with modular design and fine-grained plugin design with training flow abstraction: Modularity provides users with the flexibility to extend to new datasets and models with application-specific plugins; Training flow abstraction allows users to customize federated algorithms by replacing specific training stages in the server and client modules. 






\subsection{API}

Table \ref{tab:interfaces} summarizes the most important APIs of EasyFL. We introduce below these APIs in three categories (initialization, registration, and execution), and illustrate their usage scenarios.

\begin{table}[t]
\caption{EasyFL has three categories of APIs: initialization, registration, and execution.}
\label{tab:interfaces}
\begin{center}
\begin{tabular}{ll}
\toprule
Name & Description \\
\midrule
init(configs) & Initialize EasyFL with configurations\\
register\_dataset(train, test) & Register an external dataset\\
register\_model(model) & Register an external model\\
register\_server(server) & Register a customized server\\
register\_client(client) & Register a customized client\\
run(callback) & Start training with an optional callback\\
start\_server(args) & Start server service for remote training\\
start\_client(args) & Start client service for remote training\\
\bottomrule
\end{tabular}
\end{center}
\end{table}


\texttt{init(configs):} Initialize EasyFL with provided configurations (\texttt{configs}) or default configurations if not specified. It instructs the simulation manager to coordinate with the data manager to set up the simulation environment for experiments. Besides, these configurations determine the training hardware and hyperparameters for both experimental and production scenarios.

\texttt{register\_<module>:} Register customized modules to the system. EasyFL supports the registration of customized datasets, models, server, and client, replacing the default modules in FL training. In the experimental phase, users can register newly developed algorithms to understand their performance. In the production phase, users can use it to adapt to real-world datasets. 

\texttt{run, start\_<server/client>:} The APIs are commands to trigger execution. \texttt{run(callback)} starts FL using standalone training or distributed training, with an optional \texttt{callback} to define execution after training is completed. \texttt{start\_server} and \texttt{start\_client} start the server and client services to communicate remotely with \texttt{args} variables for configurations specific to remote training, such as the endpoint addresses. 



\begin{lstlisting}[
    language=Python, 
    label={code:example}, 
    caption=Usage examples of EasyFL.
]
# --- Example 1: Quick start ---
configs = {"model": "resnet18"} #optional
easyfl.init(configs) #initialization
easyfl.run() #start training

# --- Example 2: Remote training ---
# Start customized remote server
easyfl.register_server(NewServer) #optional
easyfl.init() #use default configs
easyfl.start_server(args) #start service
# Start customized client server
easyfl.register_client(NewClient) #optional
easyfl.init() #use default configs
easyfl.start_client(args) #start service
\end{lstlisting}

These APIs empower users with various levels of expertise to conduct Fl experiments or build FL prototypes with minimal coding, as shown in Listing \ref{code:example}. For example, beginners can quickly start FL with no more than three lines of code (Example 1); To build prototypes, experienced researchers can customize federated algorithms in the server and client modules by registering new server and client, and start the services to conduct remote training (Example 2). We further explain how to achieve low-code implementations of the server and client in Section \ref{sec:workflow-abstraction}.

\section{Ease of Experimentation and Development}
\label{sec:easy-development}


To facilitate ease of experimentation and development, EasyFL provides heterogeneity simulation for fast simulation setup, training flow abstraction for minimum coding, and comprehensive tracking for easy results analysis. 



\subsection{Heterogeneity Simulation}
\label{sec:hetero-simulation} 

To minimize researchers' hassles in setting up the simulation environment to start experiments, EasyFL provides various methods to simulate statistical and system heterogeneity using the simulation manager and the data manager. Users can easily change configurations in initialization to customize simulation methods.

\textbf{Statistical Heterogeneity} To simulate statistical heterogeneity, which is closely related with datasets, EasyFL includes three most commonly used datasets in FL research: FEMNIST \cite{leaf, cohen2017emnist}, Shakespeare \cite{shakespeare2007shakespeare, fedavg}, and CIFAR-10 \cite{cifar2009}, the statistics of which are shown in Table \ref{tab:datasets-models}. FEMNIST and Shakespeare datasets simulate the non-IID and unbalanced data with realistic scenarios as described in \cite{leaf}. CIFAR-10 dataset can be flexibly constructed to different numbers of clients with unbalanced data simulation and two different types of non-IID simulation: (1) dividing the dataset by Dirichlet process $Dir(\alpha)$ \cite{fedma}; (2) dividing the dataset by class, each client containing $N$ out of 10 classes \cite{zhao2018non-iid}. EasyFL provides flexibility for researchers to simulate various numbers of clients, applying non-IID data simulation and unbalanced data simulation, separately or in combination. 

Besides these out-of-the-box datasets, researchers can also employ \texttt{register\_dataset} API to easily integrate new datasets. For datasets that naturally simulate statistical heterogeneity, such as datasets provided in LEAF benchmark \cite{leaf}, EasyFL supports researchers to start training once the datasets are adapted into the system. For datasets that are not simulated yet, such as ImageNet dataset \cite{imagenet_cvpr09}, EasyFL provides functionalities to partition the datasets for non-IID and unbalanced data simulations. These three datasets provide a good starting point, we will keep extending to provide more datasets for simulation.

\textbf{System Heterogeneity} Although system heterogeneity is associated with hardware resources, EasyFL simulates it in a lightweight and realistic manner. System heterogeneity is caused by varieties of hardware and network connectivity of clients, resulting in varied training times among clients and network transmission times between the server and clients. As a result, the server receives model updates from clients at different times --- the slower ones are the stragglers. A significant system heterogeneity simulation should reveal these stragglers with varied time between server model distribution and client model uploads. 

We simulate the time differences using training speed variances of mobile devices from AI-Benchmark \cite{ignatov2018aibenchmark}. We first calculate the training speed ratio of different mobile devices. Then we assign each client a type of mobile device based on speed ratio. In each round of training, clients not only execute the training but also wait for the time proportional to their speed ratios before uploading updates to the central server. Besides, EasyFL can simulate networking conditions (e.g., latency) for communication with an isolated environment provided by containerization (Section \ref{sec:deployment}). As such, EasyFL simulates the system heterogeneity with stragglers. 


\subsection{Training Flow Abstraction}
\label{sec:workflow-abstraction}

\begin{figure}[ht]
    \centering
    \includegraphics[width=0.47\textwidth]{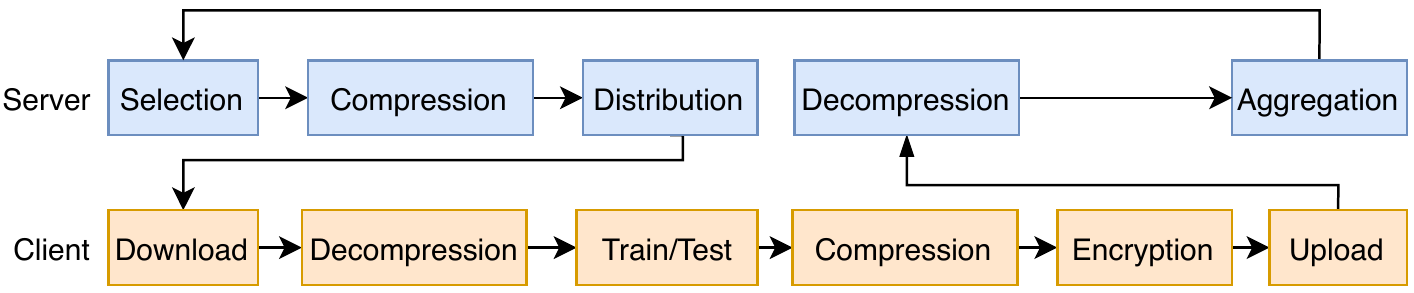}
    \caption{Training flow abstraction divides the training process in the server and client modules into multiple stages.}
    \label{fig:training-flow}
\end{figure}

EasyFL supports fine-grained plugin design by abstracting the FL training flow in the server and client modules to multiple stages as shown in Fig. \ref{fig:training-flow}. It provides researchers with the flexibility to customize any of these stages when developing new FL algorithms.


Each round of training or testing begins from the \textit{selection stage}, which employs strategies to select a fraction of participated clients in the server. The \textit{compression stage} implements data compression algorithms to reduce communication costs and the \textit{decompression stage} decompresses the data. The \textit{distribution stage} transmits data from the server to the clients. Clients download the data in the \textit{download stage} and execute training or testing after decompression. The \textit{encryption stage} encrypts the updates before clients upload them to the server in the \textit{upload stage}. At the end of each round, the server aggregates these updates in the \textit{aggregation stage}. 

We propose training flow abstraction based on analysis of FL open problems \cite{kairouz2019fl-advances-open} and emerging solutions in top publications, as shown in Table \ref{tab:publications}. We analyzed 33 papers from recent top publications: 10 out of 33 ($\sim$30\%) publications innovate new algorithms by changing only one stage of the training flow; The majority ($\sim$57\%) change only two stages of the training flow. However, implementations of these studies mostly need to write the whole FL training process. In contrast, training flow abstraction allows researchers to focus on innovating algorithms, without re-implementing the whole training process. The stages in both the server and client modules serve as templates for the standard FL training process. Users can inherit them and replace one or more stages to implement new algorithms. For example, researchers who focus on improving communication efficiency can develop new compression algorithms to replace the \textit{compression} related stages. We integrate a compression algorithm \cite{sattler2019stc} as an example with around 80 lines of code, whereas the released implementation requires several hundred lines of code. 


Moreover, EasyFL encourages users to publish their new federated algorithms as plugins to the community. Others can easily use these plugins to reproduce results and further develop on them with minimal coding. We believe that EasyFL has a great potential to grow and become an open-source community to facilitate the reproducibility of FL research.


\begin{algorithm}[t]
    \caption{Greedy Allocation with Adaptive Profiling}
    \label{alg:greedy}
    \begin{algorithmic}[1]
    \State {\bfseries Input:} number of GPUs $M$; a set of N clients $C_N$; whether client is profiled $c.profiled$; default client training time $c.time$ = $t$; update momentum $m$; 
    
    
    \For{each training round r=0 to R-1}
        \State $C_K$ = sort(random K clients of $C_N$) by $c.time$ in desc
        \State $G_M$ = initialize $M$ empty lists
        \State $T_M$ = initialize a list of $M$ 0s
        \For{each client $c \in C_K$}
            \If {not $c.profiled$}
                \State $c.time$ = $t$
            \EndIf
            \State index = \texttt{argmin} ($T_M$)
            \State $T_M$[index] += $c.time$
            \State $G_M$[index].append($c$)
        \EndFor
        \State $t$ = ADAPTIVE\_PROFILING ($G_M$, $t$)
    \EndFor
    
    
    \Function{Adaptive\_Profiling}{$G_M$, $t$}
        \For{ each group $G \in G_M$ {\bfseries in parallel} in one GPU }
            \For{ each client $c \in G$} 
                \If {not $c.profiled$} 
                    \State $c.profiled$ = True
                    \State $c.time$ = training time of client $c$
                \EndIf
            \EndFor
        \EndFor
        
        \State $C_K$ = aggregate clients $G_M$ from $M$ GPUs
        \State $t_{avg}$ = average(sum($c.time$ {\bfseries for each} client $c \in C_K$))
        \State $t_{avg}$ = $t_{avg}$ * $m$ + $t$ * (1 - $m$)
        
        \State \textbf{return} $t_{avg}$
    \EndFunction
    \end{algorithmic}
\end{algorithm}

\subsection{Comprehensive Tracking}
\label{sec:tracking}

To support comprehensive and easy analysis of FL training results, EasyFL provides a powerful tracking manager. The tracking manager is specially designed to collect and store three levels of FL metrics: training task metrics, round metrics of a task, and client metrics of a round, equipping researchers with in-depth analysis with comprehensive details. Task metrics include information about the whole training such as configurations and hyperparameters. Round metrics record the metrics in the server for each round of training and testing, such as accuracy, communication cost, and training time. Client metrics contain training and testing metrics of selected clients.

EasyFL supports two forms of tracking for diverse training methods: (1) \textit{Local tracking} logs metrics to local storage, which is efficient for standalone and distributed training; (2) \textit{Remote tracking} starts a tracking service to collect metrics via API calls, required by remote training. 

The tracking manager provides command-line tools to query the metrics. It also exposes simple APIs for gPRC calls and HTTP requests, which are extensible to build a visualization dashboard and real-time performance monitoring.

\section{Distributed Training Optimization}
\label{sec:distributed-training-optim}

EasyFL accelerates distributed training under heterogeneous simulations and resource constraints in the distribution manager. Although FL clients are deployed to edge devices such as mobile phones in real-world scenarios, researchers \textit{simulate FL experiments mainly on GPUs} for faster iteration of algorithm development \cite{leaf, tff, fate, fedprox, lai2020oort, tifl}. To further speed up algorithm development iterations in the experimental phase, we propose distributed training optimization. 

EasyFL enables the distributed training optimization with only one line change in configurations. It allows training with the number of selected clients larger than available hardware resources by allocating multiple clients into one hardware to conduct training. The allocation is not trivial because the training time of clients could vary dramatically under system heterogeneity or unbalanced data. We formulate the problem with the goal that the overall training time is shortest.

\textbf{Problem Formulation} We assume that the training times of all clients are known and describe the problem as following:
Given $M$ GPUs $\{G_1, G_2, \cdots, G_M\}$ and a set of training time of $N$ clients $S_t = \{t_1, t_2, \cdots, t_N\}$, where $N \geq M$, find the optimal way of allocating clients to GPUs such that the maximum training time across all GPUs is minimized. Basically, we aim to partition $S_t$ into $M$ mutually disjoint subsets $S_1, S_2, \cdots, S_M$ such that the total training time

\begin{equation}
  \label{eqn:gpu-allocate}
  T = \min_{i \in [1, M]} \max_{x \in S_i} \sum x,  
\end{equation}
where $\sum_{x \in S_i} x$ is the total training time in GPU $G_i$.


\textbf{Greedy Allocation with Adaptive Profiling} The problem is a variant of multiprocessor scheduling problem \cite{graham1969bounds, johnson1974worst}, which is an NP-hard optimization problem. We adopt the greedy algorithm (Longest Processing Time algorithm) to solve the problem: sort the clients in descending order by training time and allocate the slowest client to the GPU with the shortest total time.

However, the training time for each client is unknown at the beginning of the training. We use an adaptive strategy to update the training times of clients instead of profiling all the clients at the start of training like \cite{tifl, flash} because profiling would consume intolerable time when total clients are large. 
We summarize the algorithm of greedy allocation and adaptive profiling in Algorithm \ref{alg:greedy}, which we termed as Greedy Allocation with Adaptive Profiling (GreedyAda). GreedyAda updates the training time of selected clients and marks them as profiled after they complete each round of training. It then updates the training time of not-profiled clients with the \textit{moving average} of selected clients' training time and the preset default training time $t$. The hyperparameter, default client training time, is to support efficient training during profiling by mitigating extreme long or short training time of clients. Users can either set the default client training time by the estimated average training time if they are familiar with the task, or set the update momentum $m=1$ to disable it if they are uncertain about the time. The latter manner may occasionally lead to the extreme case in \textit{only one round} due to the randomness of client selection, but it does not have much impact when the majority of clients are profiled after several training rounds.

\section{Seamless and Scalable Deployment}
\label{sec:deployment}

To achieve faster iterations between the experimental and production phase, EasyFL supports seamless and scalable deployment. We first introduce remote communication that is the foundation for remote training in production. Next, we introduce containerization and service discovery mechanisms for seamless and scalable deployment. 




\begin{figure}[h]
  \centering
  \hfill
  \subfigure[]{\label{fig:containerization}\includegraphics[width=30mm]{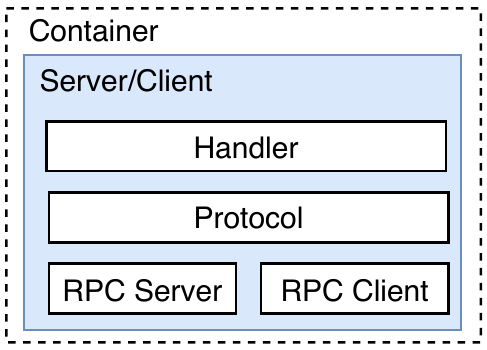}}
  \hfill
  \subfigure[]{\label{fig:service-discovery}\includegraphics[width=50mm]{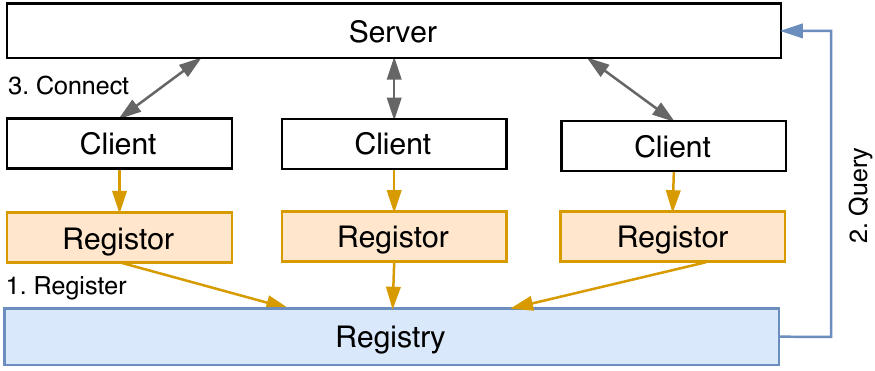}}
  \hfill
  \caption{Illustration of the architectures that support seamless and scalable deployment: (a) remote communication; (b) service discovery.}
  \label{fig:deployment}
\end{figure}

\textbf{Remote Communication} In production, remote communication supports message (model parameters or gradients) transmission between the server and clients, which is the most crucial component as the server and clients may locate in different locations.
Fig. \ref{fig:containerization} shows the three-tier architecture of the server and the client that supports remote communication. At the distribution stage of the training flow, the server serializes the operation request (training/testing) with \texttt{Protocol} and distributes them to parallel-running clients with \texttt{RPC Client}. These requests are asynchronous because clients would take a long time to execute these operations. The \texttt{RPC server} receives messages from clients after they complete execution, passing to the \texttt{Protocol}. After the \texttt{Protocol} deserializes the messages, the \texttt{Handler} processes them to streamline the training pipeline.

Since training flow abstraction (Section \ref{sec:workflow-abstraction}) decouples the training and communication, EasyFL integrates remote communication by providing alternative implementations for distribution stage and upload stage. When the system is started with \texttt{start\_server} or \texttt{start\_client} API, EasyFL switches on remote communication.

\textbf{Containerization}
Based on the remote communication, EasyFL containerizes the server, the client, and the tracking service for easy and reliable deployment to the cloud infrastructure and the edge devices, focusing on the edge devices that support containers. EasyFL containerizes using Docker \cite{docker}, which is the standard industrial containerization tool, to easily adapt to complex software dependencies in diverse computing environments of edge devices. On the one hand, containerization enables the simulation of networking conditions for system heterogeneity with simple configurations when starting the containers. On the other hand, containerization of FL further facilitates seamless deployment and builds the foundation to deploy FL services to edge computing frameworks such as AWS IoT Greengrass and KubeEdge.

\textbf{Service Discovery}
EasyFL provides a service discovery mechanism for the server to discover the clients when scaling up. Fig. \ref{fig:service-discovery} shows the architecture of service discovery, containing the \texttt{registor} to dynamically register the clients and the \texttt{registry} to store the client addresses for the server to query. The \texttt{registor} gets the addresses of clients and registers them to the registry. Since the clients are unaware of the container environment they are running, they must rely on a third-party service (the registor) to fetch their container addresses to complete registration. The \texttt{registry} stores the registered client addresses for the server to query.
EasyFL supports two service discovery methods targeting different deployment scenarios: (1) deployment using Kubernetes \cite{k8s}, which is an industrial-level container orchestration engine; (2) deployment using only Docker \cite{docker} containers. 

\begin{table}[t]
\caption{Datasets and models provided by EasyFL to simulate statistical heterogeneity. }
\label{tab:datasets-models}
\begin{center}
\begin{tabular}{llll}
\toprule
Datasets & \# of samples & \# of clients & Models \\
\midrule
FEMNIST & 805,263 & 3,550 &  CNN (2 Conv + 2 FC) \\
Shakespeare & 4,226,158 & 1,129 & RNN (2 LSTM + 1 FC) \\
CIFAR-10 & 60,000 & Flexible & ResNet18 \\

\bottomrule

\end{tabular}
\end{center}
\end{table}

\begin{figure*}[ht]
  \centering
  \subfigure[FEMNIST]{\label{fig:speed-comparison-c}\includegraphics[width=45mm]{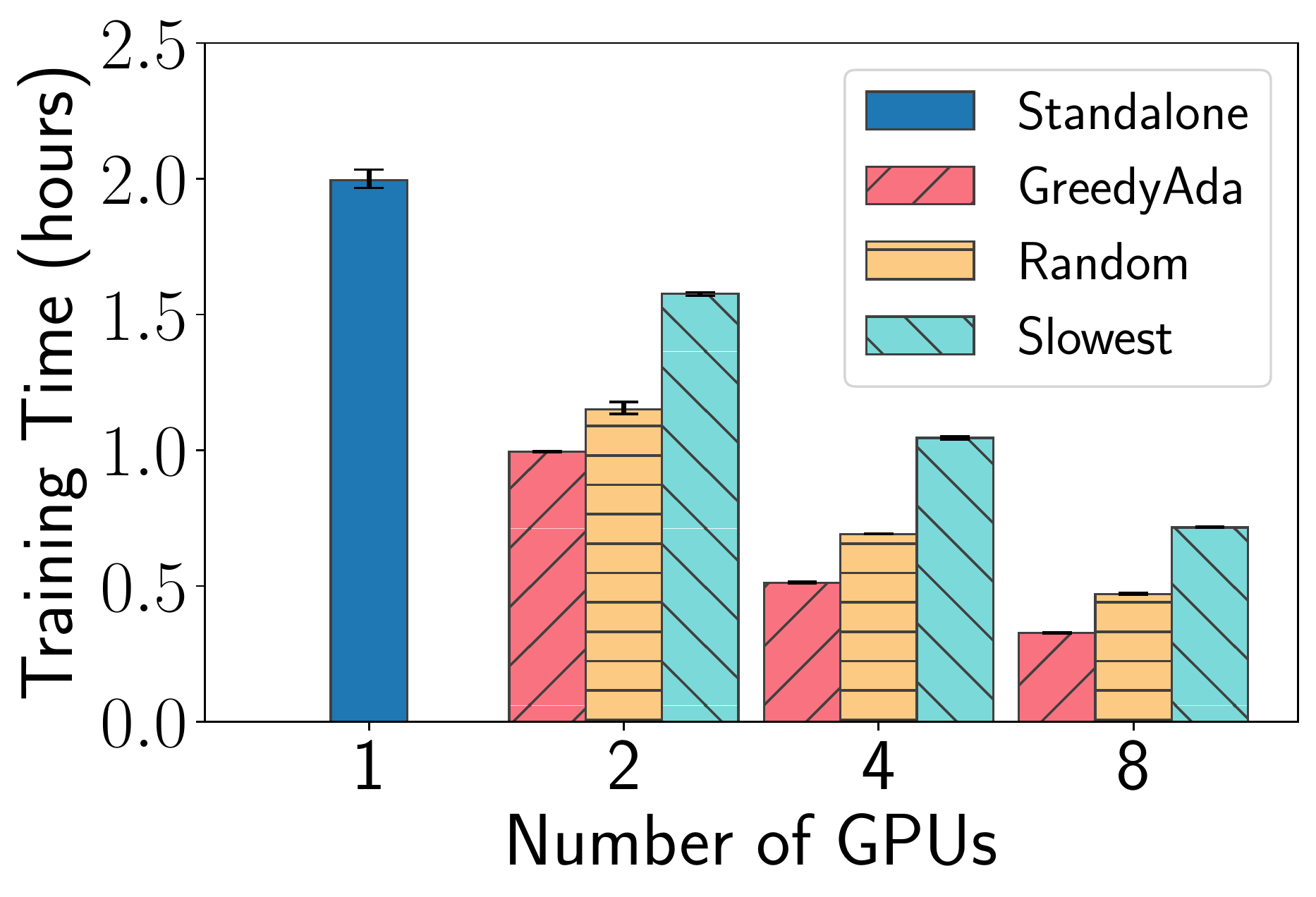}}
  \hspace{10mm}
  \subfigure[Shakespeare]{\label{fig:speed-comparison-b}\includegraphics[width=45mm]{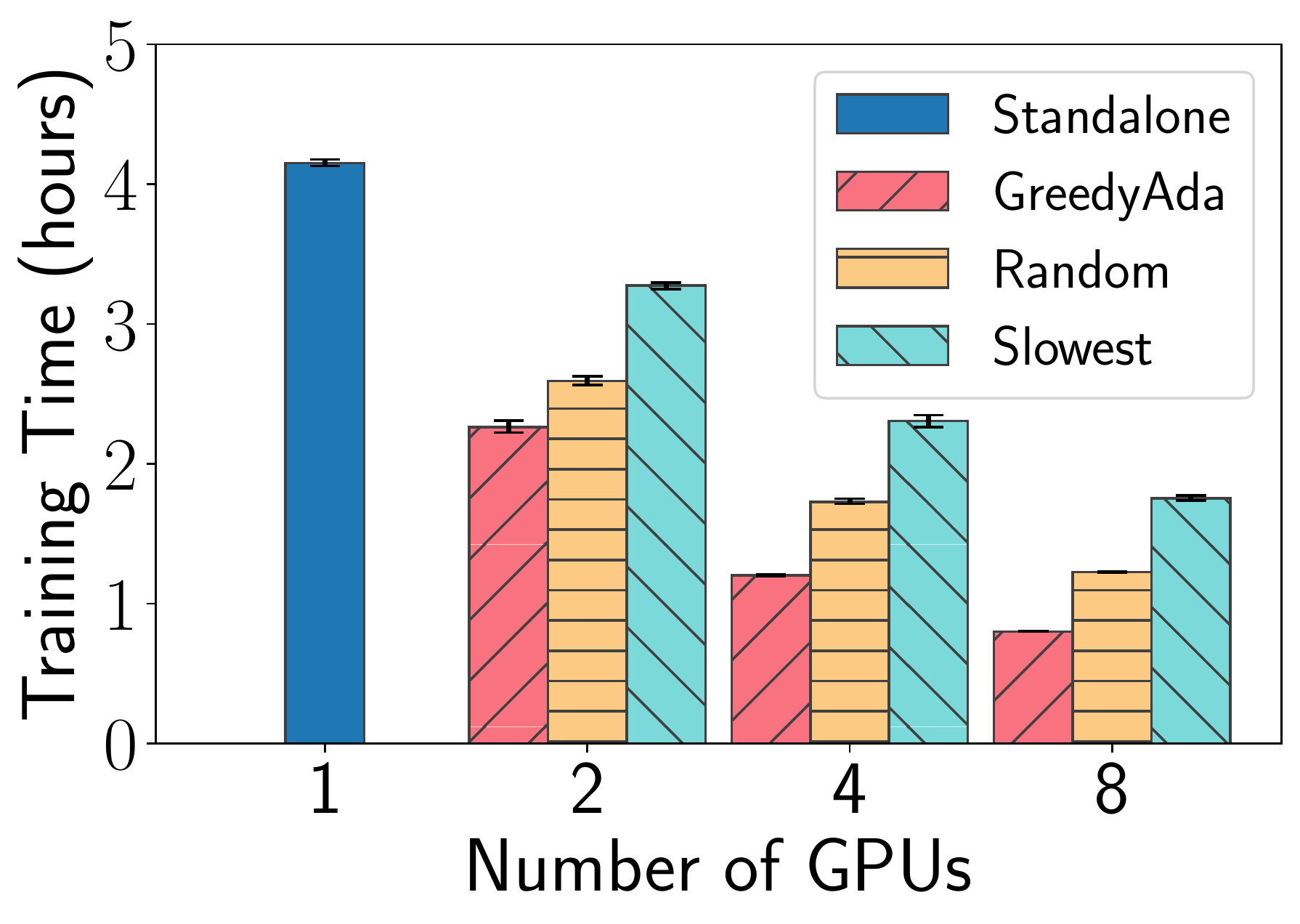}}
  \hspace{10mm}
  \subfigure[CIFAR-10]{\label{fig:speed-comparison-a}\includegraphics[width=45mm]{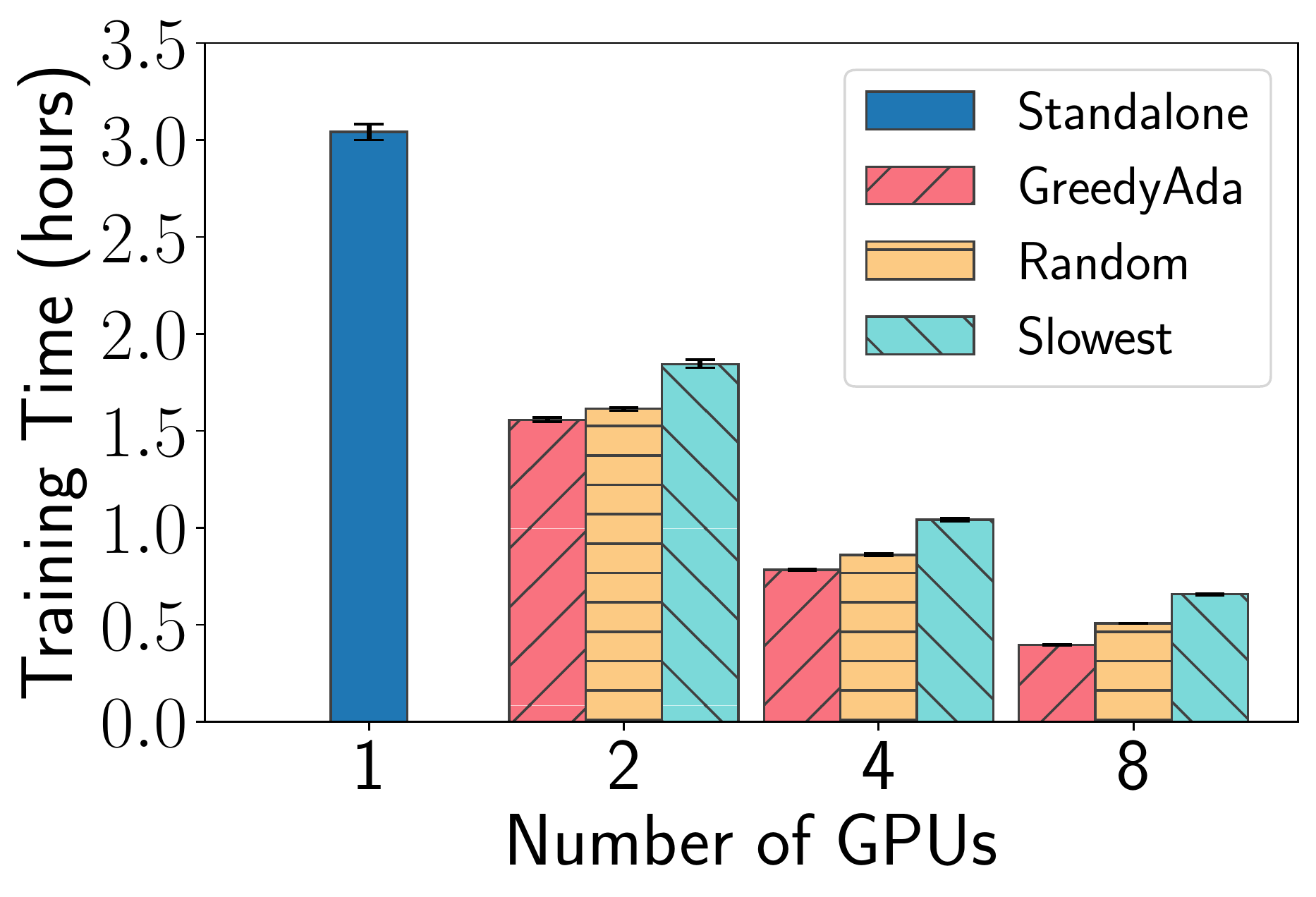}}
  \caption{Training time comparison of standalone and distributed training using three datasets. Greedy Allocation with Adaptive Profiling (GreedyAda) effectively accelerates training on all datasets.  We run the experiments with 20 selected clients per round under simulated heterogeneous scenarios.}
  \label{fig:speed-comparison}
\end{figure*}

\section{Evaluation}
\label{sec:experiments}

In this section, we start by presenting the implementation details and experimental setup. We followed by evaluating EasyFL on heterogeneity simulation, distributed training optimization, and remote training. Then we compare the lines of code to EasyFL implementations with original implementations. We end by comparing the training overhead with other FL platforms and presenting a case study of EasyFL.


\subsection{Implementation}
\label{sec:implementation}

We implement EasyFL as a standalone Python library of $\approx$ 6000 lines of code (LOC). EasyFL uses PyTorch \cite{paszke2017pytorch} as the backend for deep learning model training and testing. Remote communication in EasyFL is based on gRPC \footnote{https://grpc.io/}, a high-performance RPC framework. We also provide several other RPC methods to start and stop training. Protocol Buffers \footnote{https://developers.google.com/protocol-buffers}, a lightweight method for serializing structured data, are used to serialize remote messages. 

Targeting two different types of deployment, EasyFL provides different stacks for service discovery. For deployment using Kubernetes \cite{k8s}, EasyFL utilizes Pod, the smallest deployable unit in Kubernetes where containers run on, as the \texttt{registor}. The Service in Kubernetes, which connects client Pods and an internal DNS, serves as the \texttt{registry}. For deployment using only Docker \cite{docker} containers, EasyFL uses etcd \footnote{https://etcd.io/} as \texttt{registry} because it is a reliable and consistent key-value store for distributed systems. It provides docker image for docker-gen \footnote{https://github.com/jwilder/docker-gen}, which serves as \texttt{registor} to get the container metadata including IP addresses.

\begin{table}[t]
\caption{Accuracy comparison of IID and non-IID simulations using EasyFL. Different non-IID data partition methods lead to different degrees of performance degradation.}
\label{tab:statisitcal-heterogeneity-impact}
\begin{center}
\begin{tabular}{llcc}
\toprule
Datasets & Non-IID accuracy & IID accuracy & Accuracy gap \\
\midrule
FEMNIST & 78.12\% & 79.85\% &  1.73\% \\
Shakespeare & 46.15\% & 50.33\% & 4.18\% \\
CIFAR-10 & 93.63\% (dir) & 94.91\% & 1.28\% \\
CIFAR-10 & 89.06\% (class(3)) & 94.91\% & 5.85\% \\
CIFAR-10 & 73.66\% (class(2)) & 94.91\% & 21.25\% \\

\bottomrule
\end{tabular}
\end{center}
\end{table}

\begin{figure}[t]
  \centering
  \subfigure[]{\label{fig:hetero-impact-cifar-a}\includegraphics[width=25mm]{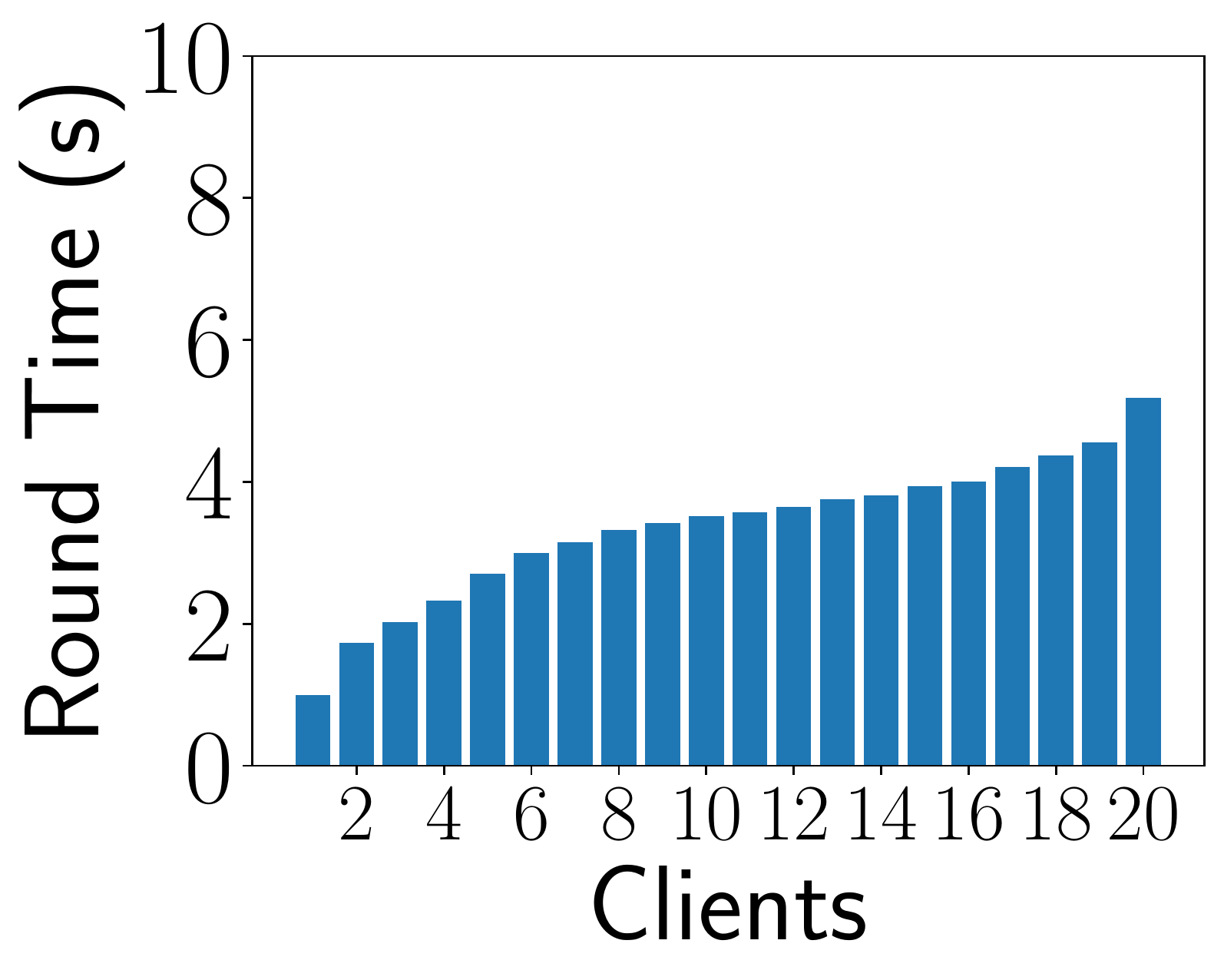}}
  \subfigure[]{\label{fig:hetero-impact-cifar-b}\includegraphics[width=25mm]{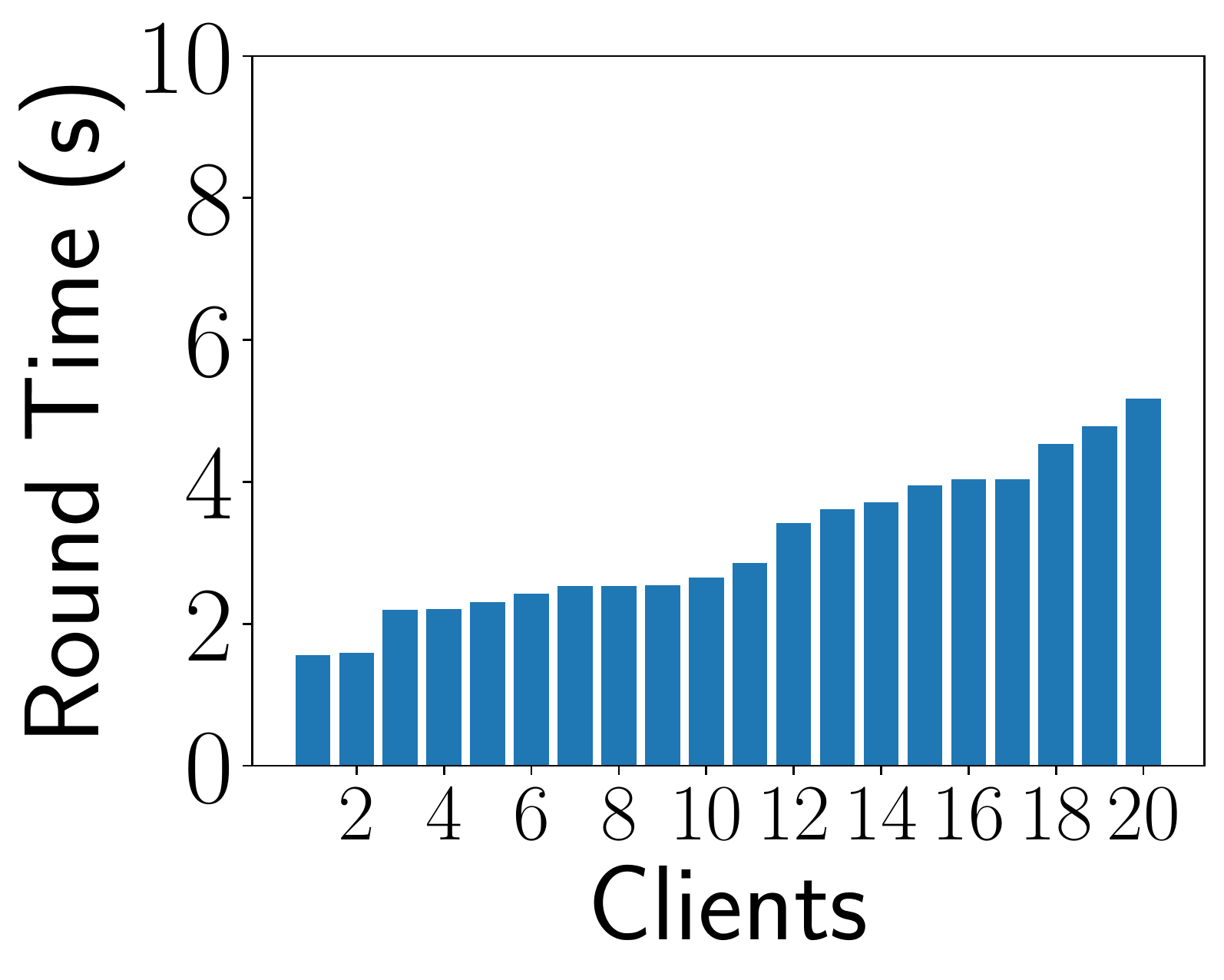}}
  \subfigure[]{\label{fig:hetero-impact-cifar-c}\includegraphics[width=25mm]{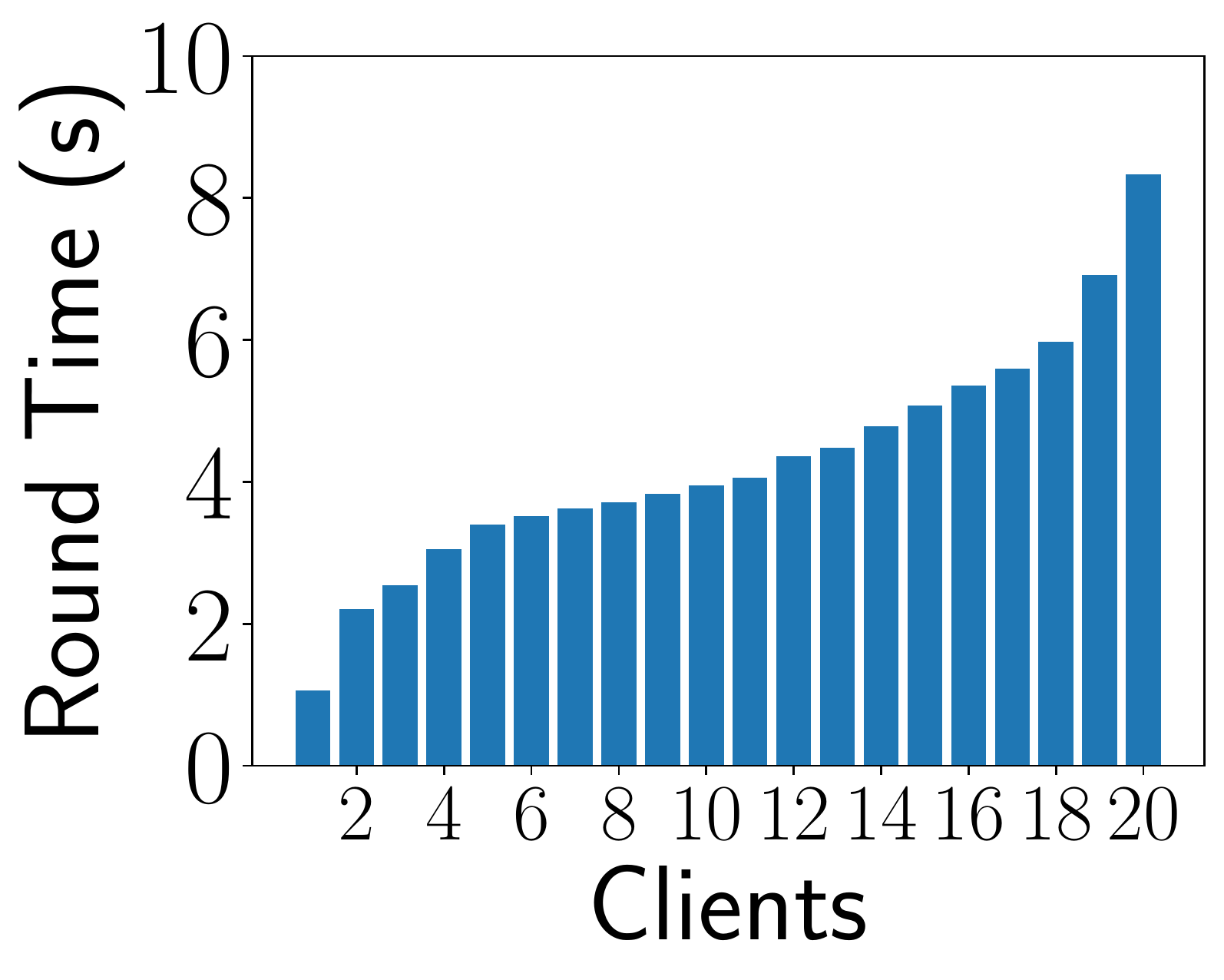}}
  \caption{Impact of heterogeneity simulation on training time per round of sampled 20 clients on CIFAR-10 dataset: (a) unbalanced data, (b) system heterogeneity, and (c) combined simulation of (a) and (b). All simulations cause training time variances, especially the combination method.}
  \label{fig:hetero-impact-cifar}
\end{figure}

\subsection{Experimental Setup}


\textbf{Datasets and Models} We use datasets and models provided by EasyFL shown in Table \ref{tab:datasets-models}. We evaluate the performance under IID and non-IID partition of these datasets.


\textbf{Algorithm and Hyperparameters} EasyFL provides FedAvg \cite{fedavg} as the standard algorithm. By default, we set local epoch to be 10 ($E = 10$) and batch size to be 64 ($B = 64$) for all the experiments. We use SGD as the optimizer and tune the learning rate for each dataset. For different experiments, we select different numbers of clients to participate in the training. More experiment settings are provided in Appendix \ref{sec:appx-experiment-setting}.

\textbf{Evaluation Metrics} We evaluate EasyFL both qualitatively and quantitatively. On the one hand, we provide qualitative results of lines of code (LOC) to implement FL applications. On the other hand, we use the evaluation metrics provided by the tracking manager for quantitative results: model accuracy, total training time, processing time each round (round time), and communication cost. To reduce the impact of hardware instability during training or testing, we calculate the round time $T_{round}$ by averaging end-to-end processing time $T_{total}$ of $R$ rounds, $T_{round} = \frac{T_{total}}{R}$. 

\textbf{Experiment Environment} All simulation experiments are run on one NVIDIA\textsuperscript{\textregistered} 2080Ti GPU or 64 NVIDIA\textsuperscript{\textregistered} V100 GPUs (8 GPUs per node) with CUDA 10.1 and CuDNN 7.6. We deploy and evaluate EasyFL in production on a Kubernetes cluster with three nodes, where each node has 28 Intel(R) Core(TM) i9-9940X CPUs.


\begin{figure*}[t]
  \centering
  \subfigure[]{\label{fig:scale-gpu}\includegraphics[width=46mm]{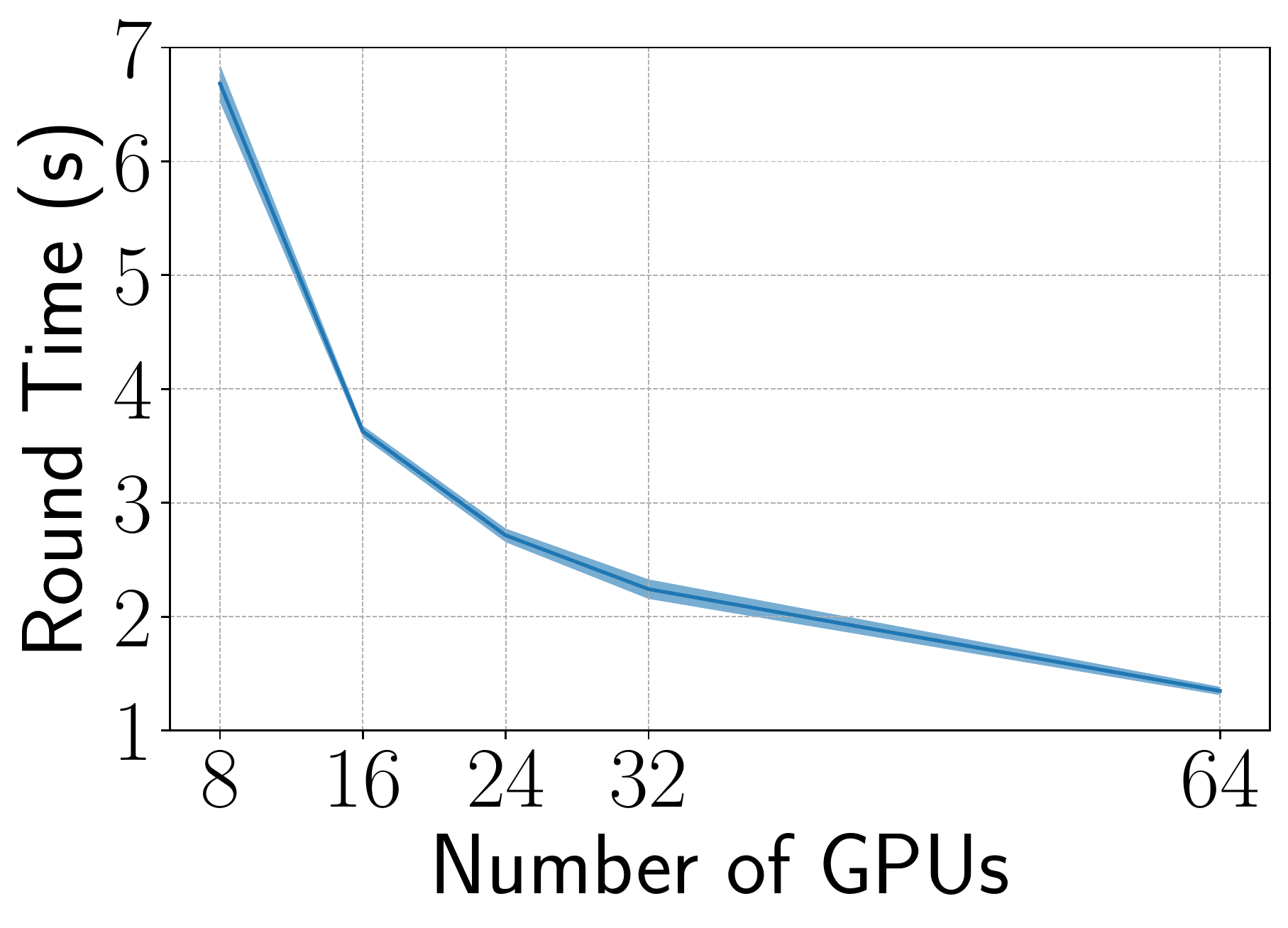}}
  \hspace{5mm}
  \subfigure[]{\label{fig:scale-data-amount}\includegraphics[width=45mm]{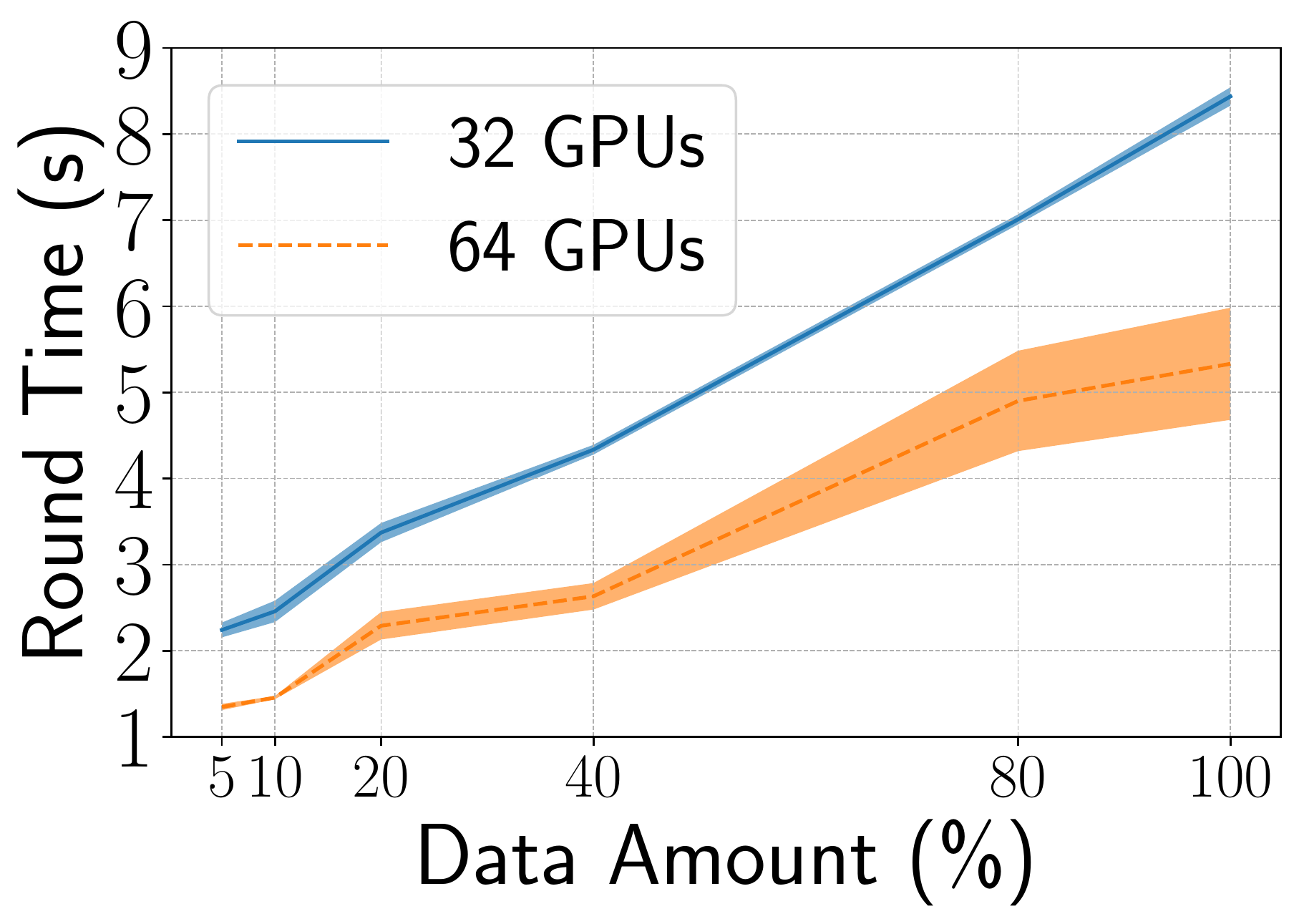}}
  \hspace{5mm}
  \subfigure[]{\label{fig:scale-data-amount-accuracies}\includegraphics[width=47mm]{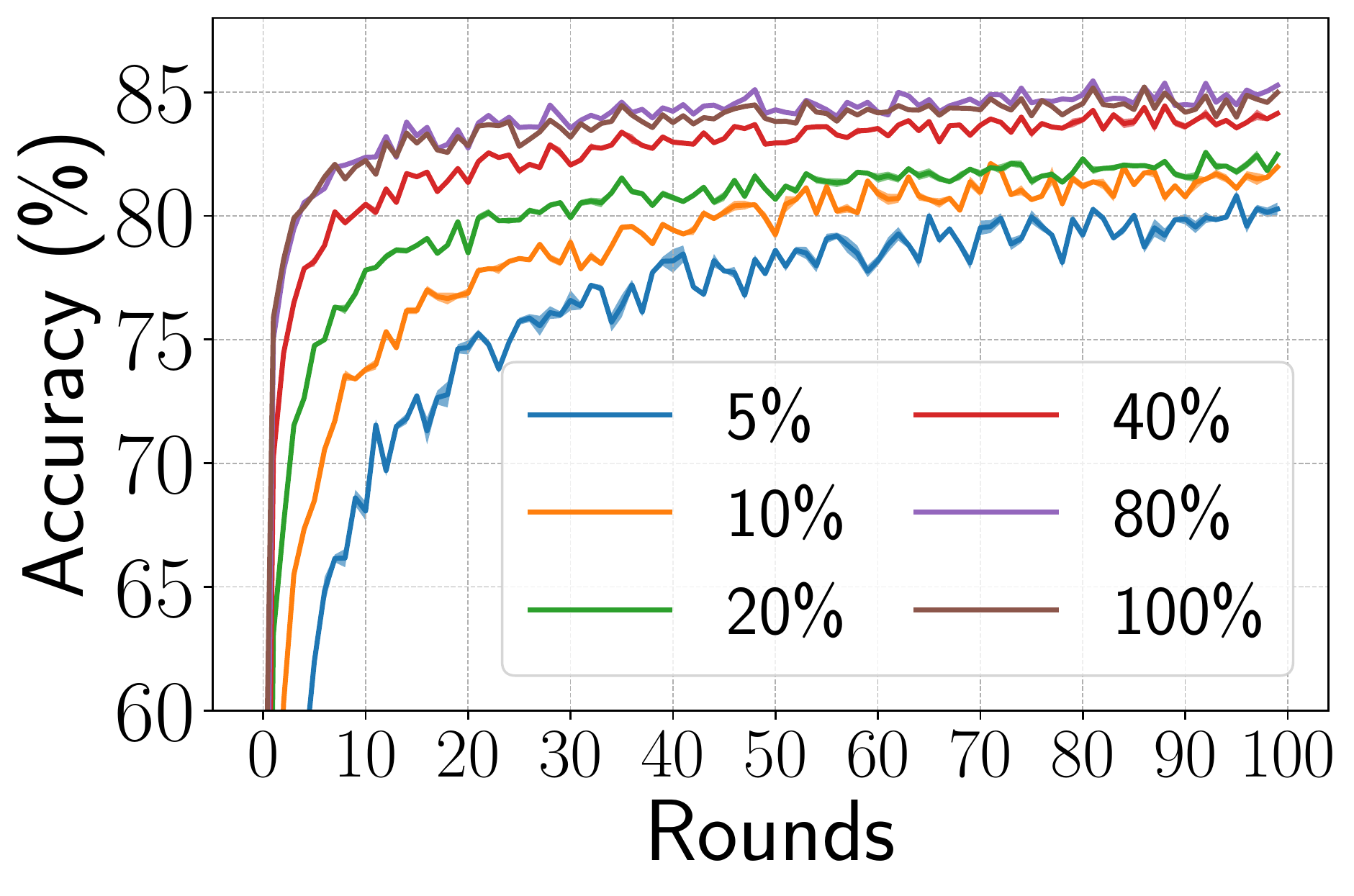}}
  \caption{Performance comparison of EasyFL on (a) round time of different numbers of GPUs, (b) round time of varied data amount, (c) accuracy of varied data amount. These results demonstrate the scalability of EasyFL: (a) the round time is effectively reduced with an increasing number of GPUs; (b) the round time increased (less than 4x) is much lower than the increased data amount (20x from 5\% to 100\%); (c) the accuracy increased from $\sim$80\% to $\sim$85\% with data amount increased from 5\% to 100\%. We run the experiments with 100 selected clients per round for 100 rounds under the IID setting.}
  \label{fig:scalability}
\end{figure*}

\subsection{Heterogeneity Simulation}
\label{sec:exp-hetero-simulation}

To show the effectiveness of heterogeneity simulation, we present benchmark results on the impact of simulated statistical heterogeneity and system heterogeneity.

\textbf{Impact of Statistical Heterogeneity on Performance} We compare the accuracy of models trained on IID and non-IID settings. For FEMNIST and Shakespeare, we simulate non-IID with realistic partition. For CIFAR-10, we simulate three levels of non-IID with increasing heterogeneity: Dirichlet process $Dir(0.5)$ (10 classes per client), three classes per client, and two classes per client. These experiments are run with 10 selected clients per round. 

Table \ref{tab:statisitcal-heterogeneity-impact} shows that the simulated statistical heterogeneity leads to performance degradation. Models trained on simulated non-IID data all perform worse than models trained on the IID data. In particular, the increasing degree of statistical heterogeneity simulated from \textit{Dir} to two classes per client causes even larger degradation, leading to the largest accuracy gap of 21.25\%. Researchers can use these results as a benchmark when they start using EasyFL and have the flexibility to customize the degree of non-IID.


\textbf{Impact of Heterogeneity on Training Time} We simulate unbalanced data by $Dir(0.5)$ and system heterogeneity using the EasyFL simulation manager. Fig. \ref{fig:hetero-impact-cifar} shows that unbalanced data, system heterogeneity, and their combination cause a huge discrepancy in training time of each round among clients in the CIFAR-10 dataset. The fastest client is four times faster than the slowest client because of unbalanced data as shown in Fig. \ref{fig:hetero-impact-cifar-a}. This gap is larger under system heterogeneity in Fig. \ref{fig:hetero-impact-cifar-b} and their combination in Fig. \ref{fig:hetero-impact-cifar-c}.
FL training normally requires hundreds of rounds, further amplifying the impact of the stragglers. We provide the results of training time per round of Shakespeare and CIFAR-10 datasets in Appendix \ref{sec:appx-experiments-results}. 

\begin{figure}[t]
  \centering
  \includegraphics[width=0.27\textwidth]{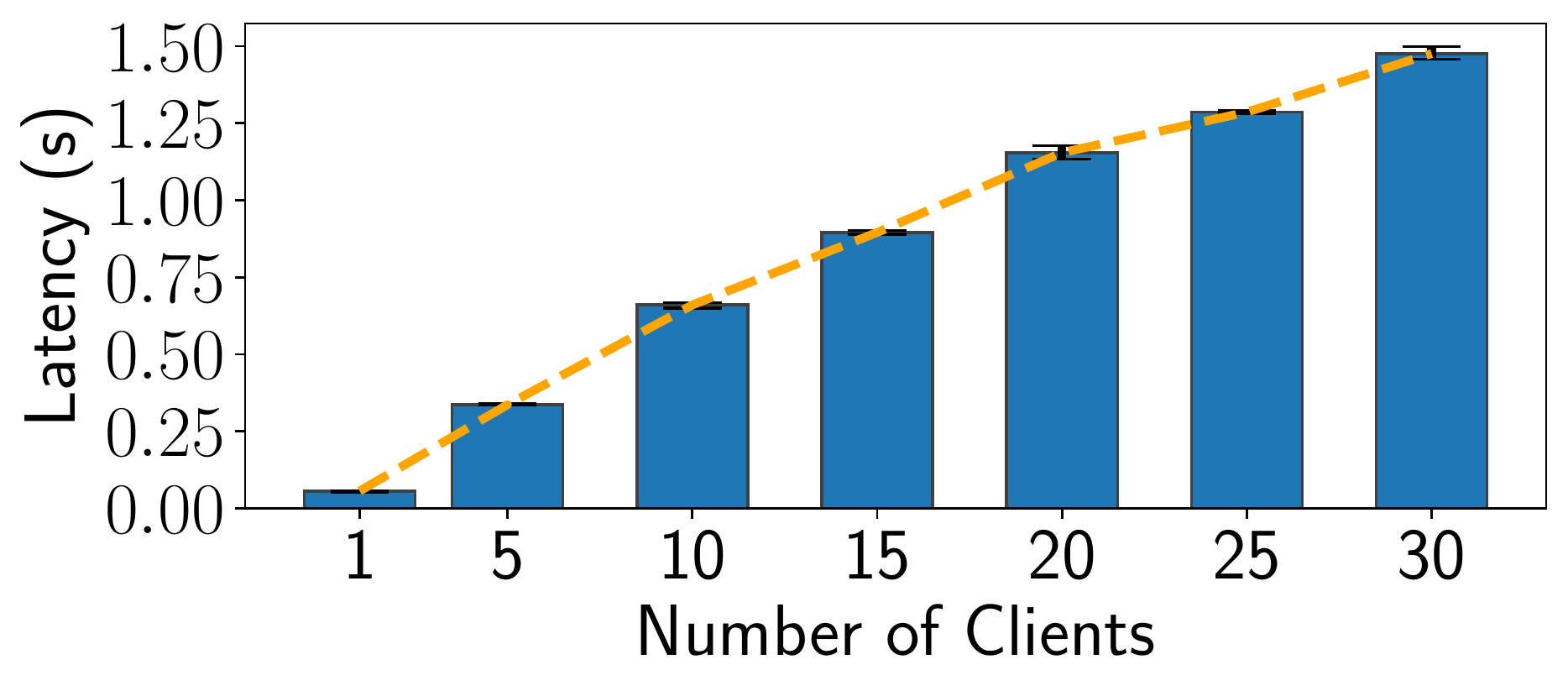}
  \caption{The distribution latency from the server to clients increases almost linearly when scaling up the number of clients in the production phase. The latency is low compared to training time.}
  \label{fig:scale-remote-training}
\end{figure}

\begingroup
\setlength{\tabcolsep}{0.3em}
\begin{table}[t]
\caption{Comparison of lines of code and round time of the original and EasyFL implementations on three different FL applications.}
\label{tab:loc-comparison}
\begin{center}
\begin{tabular}{llccccc}
\toprule
\multicolumn{1}{l}{\multirow{2}{*}{Name}} &
  \multicolumn{1}{l}{\multirow{2}{*}{Application}} &
  \multicolumn{2}{c}{Lines of Code} &
  \multicolumn{1}{c}{} &
  \multicolumn{2}{c}{Round Time} \\ \cline{3-4} \cline{6-7}
  \multicolumn{1}{c}{} & & Original & EasyFL & & Original & EasyFL\\ 
\midrule
FedProx \cite{fedprox} & Optimization & $\sim$380 & $\sim$40 & & 3.3s & 2.0s \\
STC \cite{sattler2019stc} & Compression & $\sim$560 & $\sim$80 & & 3.1s & 2.8s \\
FedReID \cite{zhuang2020fedreid} & Video surveillance & $\sim$450 & $\sim$140 & & 650.7s & 582.5s \\
\bottomrule
\end{tabular}
\end{center}
\end{table}
\endgroup

\begingroup
\setlength{\tabcolsep}{0.35em}
\begin{table*}[t]
  \caption{Comparison of training overhead (round time) and GPU usage of different FL libraries on different hardware. EasyFL is generally more efficient than LEAF and TFF.} 
  \label{tab:round-time}
  \begin{center}
  \begin{tabular}{lccccccccccccc}
  \hline
  \multicolumn{1}{l}{\multirow{2}{*}{Platform}} &
  \multicolumn{1}{l}{\multirow{2}{*}{Hardware}} &
  \multicolumn{3}{c}{FEMNIST} &
  \multicolumn{1}{c}{} &
  \multicolumn{3}{c}{Shakespeare} & 
  \multicolumn{1}{c}{} &
  \multicolumn{3}{c}{CIFAR-10} \\ \cline{3-5} \cline{7-9} \cline{11-13}
  \multicolumn{1}{c}{} & & Round Time & Util.$^a$  & Mem.$^b$  & & Round Time & Util.$^a$  & Mem.$^b$  & & Round Time & Util.$^a$  & Mem.$^b$  \\ 
  \midrule
  LEAF \cite{leaf} & 2080Ti & 7.99$\pm$0.06s (2.00x) & 12.8\% & 8\% & & 374.27$\pm$2.25s (5.71x) & 36.8\% & 5\% & & --$^c$ & -- & -- \\
  TFF \cite{tff} & 2080Ti & 5.49$\pm$0.11s (1.38x) & 13.8\% & 88\% & & 2153.91$\pm$5.74s (32.86x) & 25.5\% & 98\% & & 28.55$\pm$0.08s (1.07x) & 69.2\% & 96\% \\
  EasyFL & 2080Ti & 3.99$\pm$0.02s (1.00x) & 47.9\% & 17\% & & 65.54$\pm$1.58s (1.00x) & 53.0\% & 11\% & & 26.66$\pm$0.08s (1.00x) & 37.2\% & 91\% \\
  \hline
  LEAF \cite{leaf} & V100 & 5.63$\pm$0.14s (1.91x) & 14.2\% & 3\% & & 276.96$\pm$8.32s (4.67x) & 54.1\% & 2\% & & --$^c$ & -- & -- \\
  TFF \cite{tff} & V100 & 5.64$\pm$0.10s (1.91x) & 11.0\% & 90\% & & 1352.47$\pm$27.26s (22.78x) & 40.5\% & 95\% & & 27.97$\pm$2.42s (1.06x) & 63.43\% & 93\% \\
  EasyFL & V100 & 2.95$\pm$0.24s (1.00x)& 29.0\% & 30\% & & 59.36$\pm$0.77s (1.00x) & 62.2\% & 10\% & & 26.39$\pm$0.51s (1.00x) & 31.95\% & 31\% \\


  \hline
  \end{tabular}
  \end{center}
  \vspace{-2mm}
  \begin{tablenotes}[para]
   \item[a] Average GPU utilization.
   \item[b] Average GPU memory usage percentage.
   \item[c] LEAF does not support the CIFAR-10 dataset. 
  \end{tablenotes}
\end{table*}
\endgroup

\subsection{Distributed Training Optimization}
\label{sec:exp-distributed-training-optim}

In this section, we demonstrate the effectiveness of our proposed GreedyAda and the scalability of distributed training in EasyFL.

\textbf{Performance of GreedyAda} EasyFL accelerates training with GreedyAda (Greedy Allocation with Adaptive Profiling) under resource constraints of $M$ available GPUs and heterogeneous scenarios. The heterogeneous scenarios are simulated with the combination of unbalanced data and system heterogeneity described in Section \ref{sec:exp-hetero-simulation}. We run the experiments with 20 selected clients per round using standalone training and distributed training with different client allocation strategies: (1) GreedyAda; (2) random allocation that randomly allocates around $\frac{20}{M}$ clients to a GPU; (3) slowest allocation that allocates around $\frac{20}{M}$ slowest clients to a GPU. Fig. \ref{fig:speed-comparison} demonstrates that GreedyAda achieves the fastest training speed --- up to 1.5x faster than random allocation and up to 2.2x faster than slowest allocation --- in all datasets with different number of GPUs. 

\textbf{Scalability} We further evaluate the scalability of distributed training with different numbers of GPUs $\{8, 16, 24, 32, 64\}$ and varied amount of data $\{5\%, 10\%, 20\%, 40\%, 80\%, 100\%\}$. Data amount means the percentage of samples used training in clients. We run these experiments with 100 selected clients each round under the IID setting for 100 rounds on FEMNIST and measure the processing time per round (round time).

Fig. \ref{fig:scale-gpu} demonstrates that the round time is effectively reduced with increasing numbers of GPUs. While compared with 1.84x (optimal 2x) speedup from 8 GPUs to 16 GPUs, the speedup from 8 GPUs to 64 GPUs is only 4.96x (optimal 8x). The underutilization of hardware resources is mainly because the data amount is small (5\%) for these experiments, causing the communication overhead among GPUs to outweigh the training time when the number of GPUs is large. We further investigate it by increasing the data amount. 


Fig. \ref{fig:scale-data-amount} illustrates that the round time increases as data amount increases, either using 32 or 64 GPUs. However, the increase in round time is much smaller than the increase in data amount. In particular, the data amount increases by 20x from 5\% to 100\%, but the round time only increases less than 4x. These results suggest that when the data amount is small, the smaller number of GPUs can handle training effectively; when the data amount is large, EasyFL can scale up to the large number of GPUs for training. Besides, we also present the evaluation accuracy of different data amounts in Fig. \ref{fig:scale-data-amount-accuracies}. The accuracy increased from $\sim$80\% to $\sim$85\% with data amount increased from 5\% to 100\%.

\subsection{Remote Training}
\label{sec:exp-container}

We evaluate remote training on time to deployment and performance of training in containers.

\textbf{Deployment Time} EasyFL enables sub-linear time to deployment by leveraging the power of containerization and container orchestration engine, Kubernetes. Without containerization, the time to deployment grows linearly as we increase the number of clients, where the deployment of one client would need hours because of diverse and complex computing environments. In contrast, EasyFL only needs a one-time setup of the Docker environment and a Kubernetes cluster. It builds docker images of FL components in seconds and deploys clients to each node in Kubernetes clusters in minutes, dramatically reducing the deployment time and effort.

\textbf{Accuracy and Distribution Latency} EasyFL supports remote training to achieve the same accuracy as experiments in Table \ref{tab:statisitcal-heterogeneity-impact} under the same settings. Besides, Fig. \ref{fig:scale-remote-training} presents the distribution latency of the server when scaling up the number of clients using FEMNIST dataset. Although the distribution latency increases almost linearly using multi-threading, the latency is relatively low compared to the training time needed. To further optimize this distribution latency, we can consider replicating servers to load balance the requests.

\subsection{Applications}

EasyFL enables easy implementation of FL to a wide range of applications by allowing seamless migration from existing training codes. We demonstrate its capability by using it to implement three applications: (1) FedProx \cite{fedprox} is an optimization framework to address statistical and system heterogeneity; (2) STC \cite{sattler2019stc} is a compression framework to reduce communication cost; (3) FedReID \cite{zhuang2020fedreid} implements FL to person re-identification, an important computer vision task for video surveillance. Table \ref{tab:loc-comparison} compares lines of code (LOC) and average training time per round (round time)\footnote{The round times of FedProx and STC are evaluated on the MNIST dataset.} of EasyFL implementation with the original implementations. EasyFL greatly reduces the efforts of researchers in writing codes (3.2x to 9.5x less coding) while maintaining and even using less round time. Fewer codes result in fewer bugs and faster iterations of development. Besides, the codes are similar to non-FL implementation that researchers are familiar with. We provide more details of LOC counting and implementation of STC \cite{sattler2019stc} using EasyFL in Appendix \ref{sec:appx-application}.

\begin{figure}[t]
  \centering
  \includegraphics[width=0.27\textwidth]{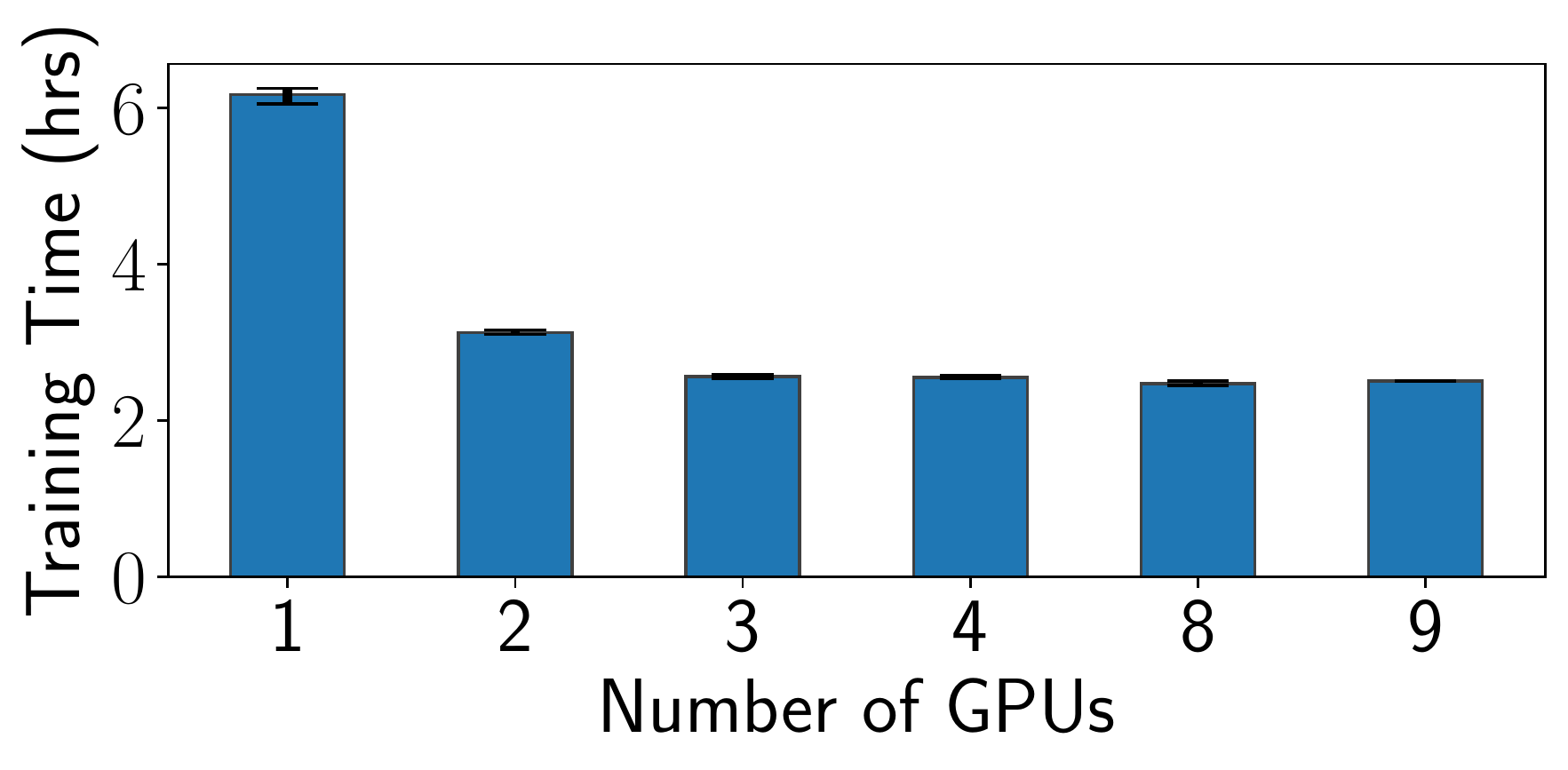}
  \caption{EasyFL achieves near-optimal training speed using 3 GPUs instead of 9 GPUs for training FedReID \cite{zhuang2020fedreid} with 9 clients, reducing the demand for hardware resources.}
  \label{fig:fedreid}
\end{figure}

\subsection{Training Overhead}

Despite that EasyFL reduces the LOC significantly with abstractions, we further investigate whether these abstractions would lead to extra training overhead by comparing with other FL libraries: LEAF \cite{leaf} and TFF \cite{tff}. We conduct the experiments with 10 selected clients per round under IID simulation for $R = 100$ rounds on FEMNIST and CIFAR-10 datasets, and $R = 20$ rounds on the Shakespeare dataset.

Table \ref{tab:round-time} compares the round time and GPU usage of EasyFL, LEAF, and TFF. We standardize the experimental environment and the versions of PyTorch and TensorFlow to conduct these experiments\footnote{These experiments are run with CUDA10.1 and CuDNN 7.6. EasyFL is run on PyTorch 1.7.0; LEAF and TFF are run on TensorFlow 2.3.2. We upgrade LEAF from TensorFlow 1.3 and use TFF v0.17.0 (latest v0.19.0) to be compatible with CUDA and CuDNN versions.}. To facilitate comparison across two deep learning frameworks (TensorFlow and PyTorch), we refer to the baseline in PyTorch paper \cite{paszke2019pytorch} for back-of-the-envelope comparison: PyTorch is 1.06 (using ResNet-50) to 2.14 times faster than TensorFlow on computer vision (CV) tasks and 1.61 times faster than TensorFlow on the natural language processing (NLP) task. EasyFL achieves similar speedups on CIFAR-10 dataset using ResNet-18, 1.38 to 2 times of speedups on FEMNIST dataset (simple CV task), and 4.67 to 32.86\footnote{TFF is slow on the Shakespeare dataset because the LSTM model is unable to use CuDNN kernel due to the FL computation.} times of speedups on Shakespear dataset (NLP task). These results show that EasyFL enables users to write less code without imposing extra system overhead. It also has better GPU utilization on FEMNIST and Shakespear datasets. Compared with TFF that tends to use all available GPU memory, LEAF and EasyFL are much more efficient in memory usage.

\subsection{Case Study}

We conduct a case study of EasyFL on developing new federated computer vision applications with FedReID \cite{zhuang2020fedreid}. FedReID implements FL to person re-identification (ReID), which is an important retrieval task in computer vision. Since FedReID is a new application that uses nine heterogeneous datasets to simulate nine clients, we adapt these datasets into EasyFL via \texttt{register\_dataset} API. As the training and testing of ReID are different from image classification, we customize the \texttt{train} and \texttt{test} in clients and integrate it with \texttt{register\_client} API. The codes for training and testing are almost the same as the ones used for normal person ReID training. With these two registrations, we then initialize EasyFL with \texttt{init} API with configurations like training rounds and learning rates. After these steps, we can start training with \texttt{run} API. 


We further leverage distributed training optimization in EasyFL to accelerate FedReID training and achieve near-optimal training speed with 6 fewer GPUs, as shown in Fig. \ref{fig:fedreid}. We train FedReID with 9 clients containing unbalanced data. The client with the largest dataset is the bottleneck in training. Instead of training with 9 GPUs by allocating each client to one of them, EasyFL saves hardware resources by achieving similar training speeds with only 3 GPUs.

Moreover, we build a prototype of FedReID using EasyFL with containerization and seamless deployment to Kubernetes. It imitates the real-world scenario that a client is deployed to an edge device, training with data collected from cameras.

\section{Conclusion}
\label{sec:conclusion}


In this paper, we propose a low-code FL platform, \textit{EasyFL}, to empower users with different levels of expertise to conduct FL experiments efficiently and deploy FL applications seamlessly with minimal coding. We achieve it with a two-tier system architecture: the interface layer contains simple APIs that shield complex system implementations for users; the system layer embraces modular design and abstracts the training flow to provide flexibility and reusability. Eight modules in the system layer provide out-of-the-box functionalities to accelerate the iteration and increase the productivity of users. EasyFL only needs 3 lines of code (LOC) to implement a vanilla FL application, which is at least 10x less than other platforms. We also implement several FL applications with 4.5x to 9.5x less LOC than the original implementations. Our evaluation demonstrates that EasyFL increases the training speed of distributed training by 1.5x. It also reduces the demand for resources and decreases the time to deployment. 

In the future, we will support out-of-the-box encryption methods in EasyFL and provide mechanisms to capture the production environment for more representative simulations. Besides, we consider a \textit{no-code} FL platform to further liberate FL for wider adoption. 









\section*{Acknowledgment}
This study is supported in part by 1) the RIE2020 Industry Alignment Fund – Industry Collaboration Projects (IAF-ICP) Funding Initiative, as well as cash and in-kind contribution from the industry partner(s); 2) the National Research Foundation, Singapore, and the Energy Market Authority, under its Energy Programme (EP Award $<$NRF2017EWT-EP003-023$>$); 3) Singapore MOE under its Tier 1 grant call, Reference number RG96/20.



\appendices

\section{Applications}
\label{sec:appx-application}

In this section, we provide more details on lines of code comparison and present an implementation example of EasyFL.

We supplement the source codes for implementations of the FL vanilla application using various frameworks
\footnote{LEAF: https://github.com/TalwalkarLab/leaf}
\footnote{TFF: https://www.tensorflow.org/federated}
\footnote{\url{PaddleFL: https://github.com/PaddlePaddle/PaddleFL/tree/master/python/paddle\_fl/paddle\_fl/examples/femnist\_demo}}
\footnote{\url{FATE: https://github.com/FederatedAI/FATE/blob/master/examples/pipeline/hetero\_nn/pipeline-hetero-nn-train-binary.py}}
\footnote{\url{PySyft: https://github.com/OpenMined/PySyft/blob/dev/examples/experimental/madhava/MNIST.ipynb}}
.


We compare EasyFL implementations with original implementations on lines of code for specific applications in Table \ref{tab:loc-comparison}, not counting the lines of the import statements. We refer to the original implementations in Github 
\footnote{STC: https://github.com/felisat/federated-learning} \footnote{FedProx: https://github.com/litian96/FedProx}
\footnote{FedReID: https://github.com/cap-ntu/FedReID}. 


\section{Experiments}
\label{sec:appx-experiments}

This section provides detailed experiment settings and two complementary experiment results.

\subsection{Experiment Settings}
\label{sec:appx-experiment-setting}

By default, we use batch size $B = 64$ and local epoch $E = 10$. We use SGD with momentum 0.9 as the optimizer with learning rates $\eta = 0.01$ for FEMNIST dataset and CIFAR-10 dataset and $\eta = 0.8$ for Shakespeare dataset.

For experiments on \textit{Impact of Statistical Heterogeneity} in Section VIII-C, we use the number of selected clients $C = 10$ and total training rounds $R = 150$ for FEMNIST dataset \cite{cohen2017emnist, leaf}, and $R=100$ for both Shakespeare \cite{shakespeare2007shakespeare, leaf} and CIFAR-10 \cite{cifar2009} datasets.


\begin{figure}[h]
  \centering
  \subfigure[]{\label{fig:hetero-impact-femnist-a}\includegraphics[width=25mm]{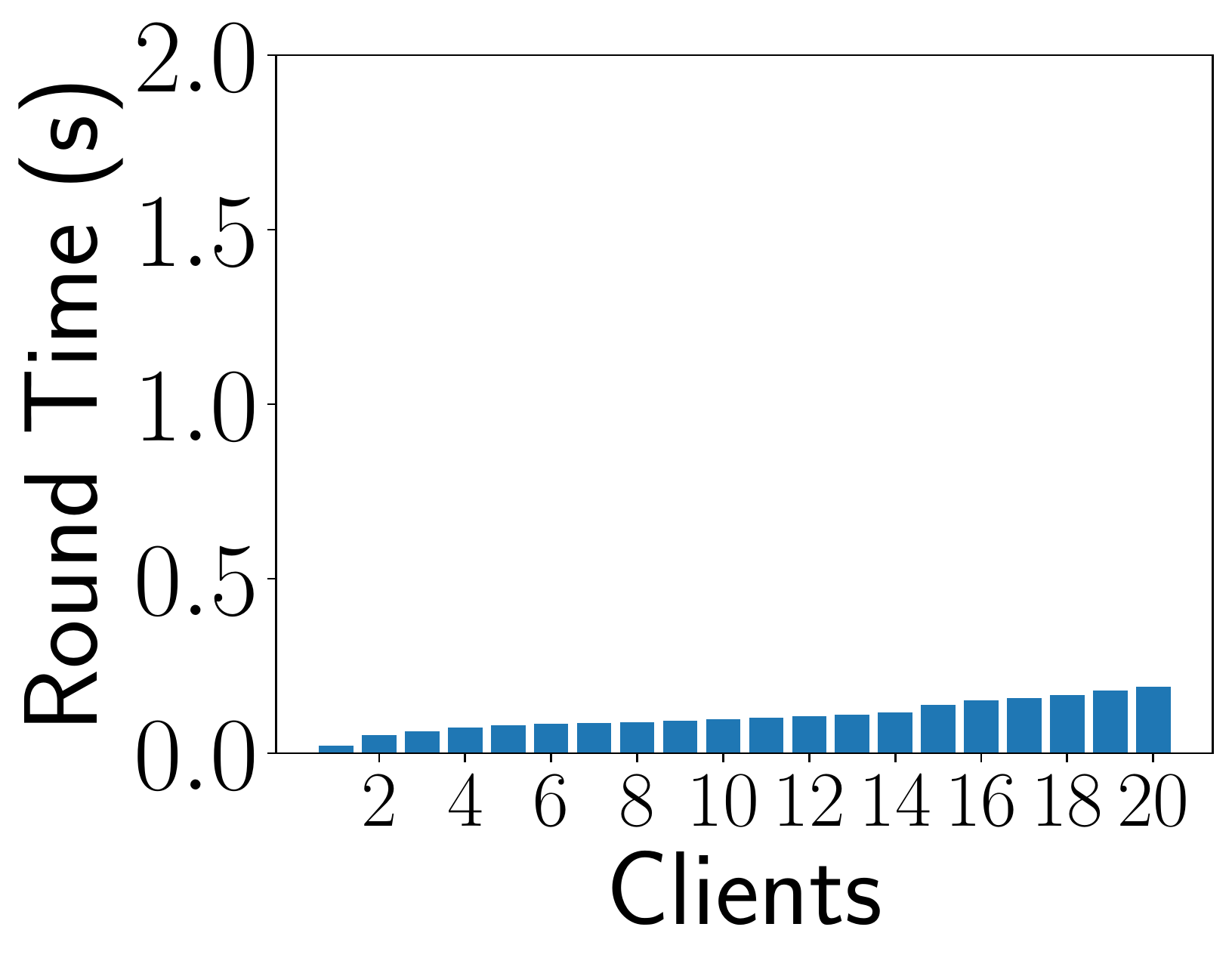}}
  \subfigure[]{\label{fig:hetero-impact-femnist-b}\includegraphics[width=25mm]{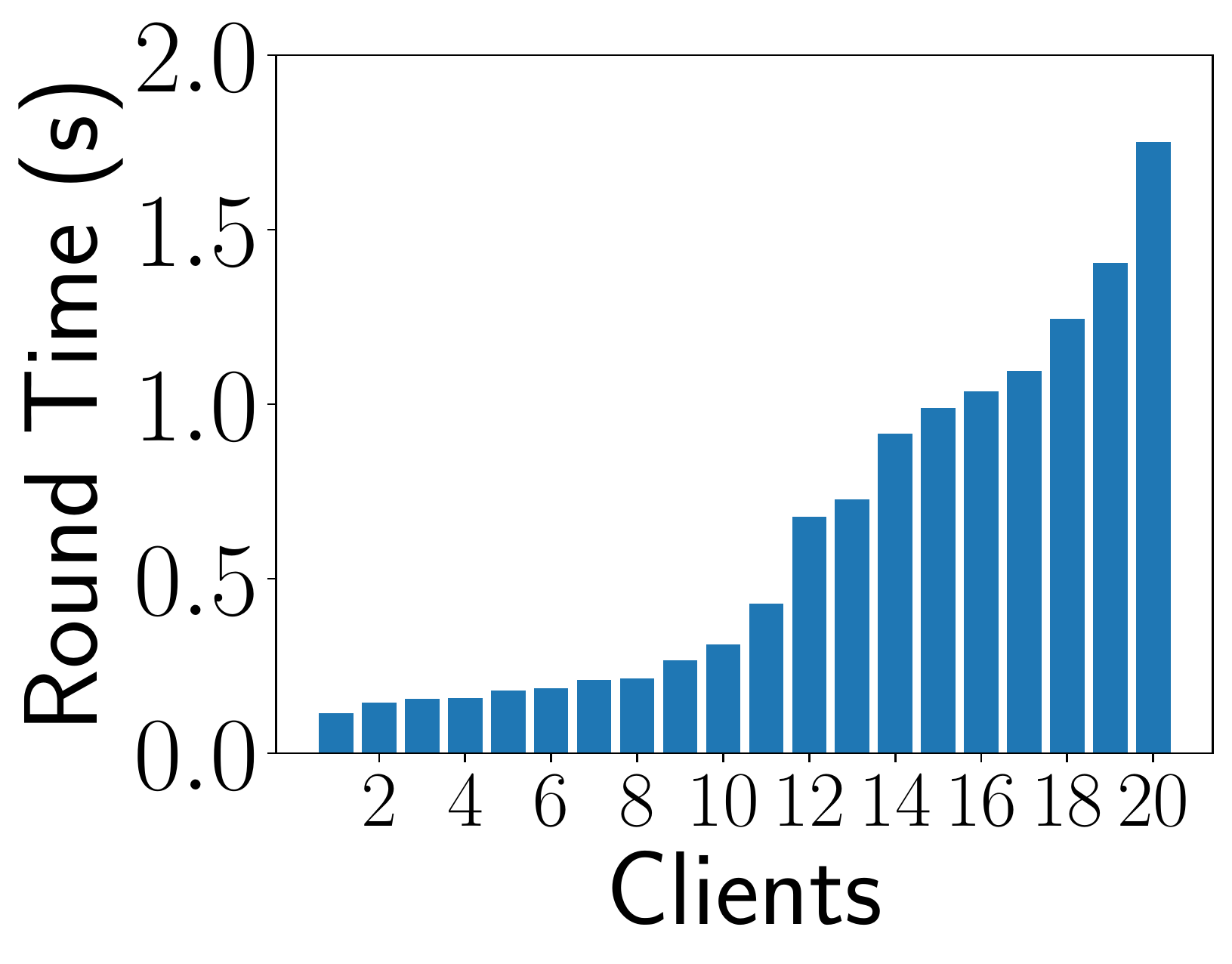}}
  \subfigure[]{\label{fig:hetero-impact-femnist-c}\includegraphics[width=25mm]{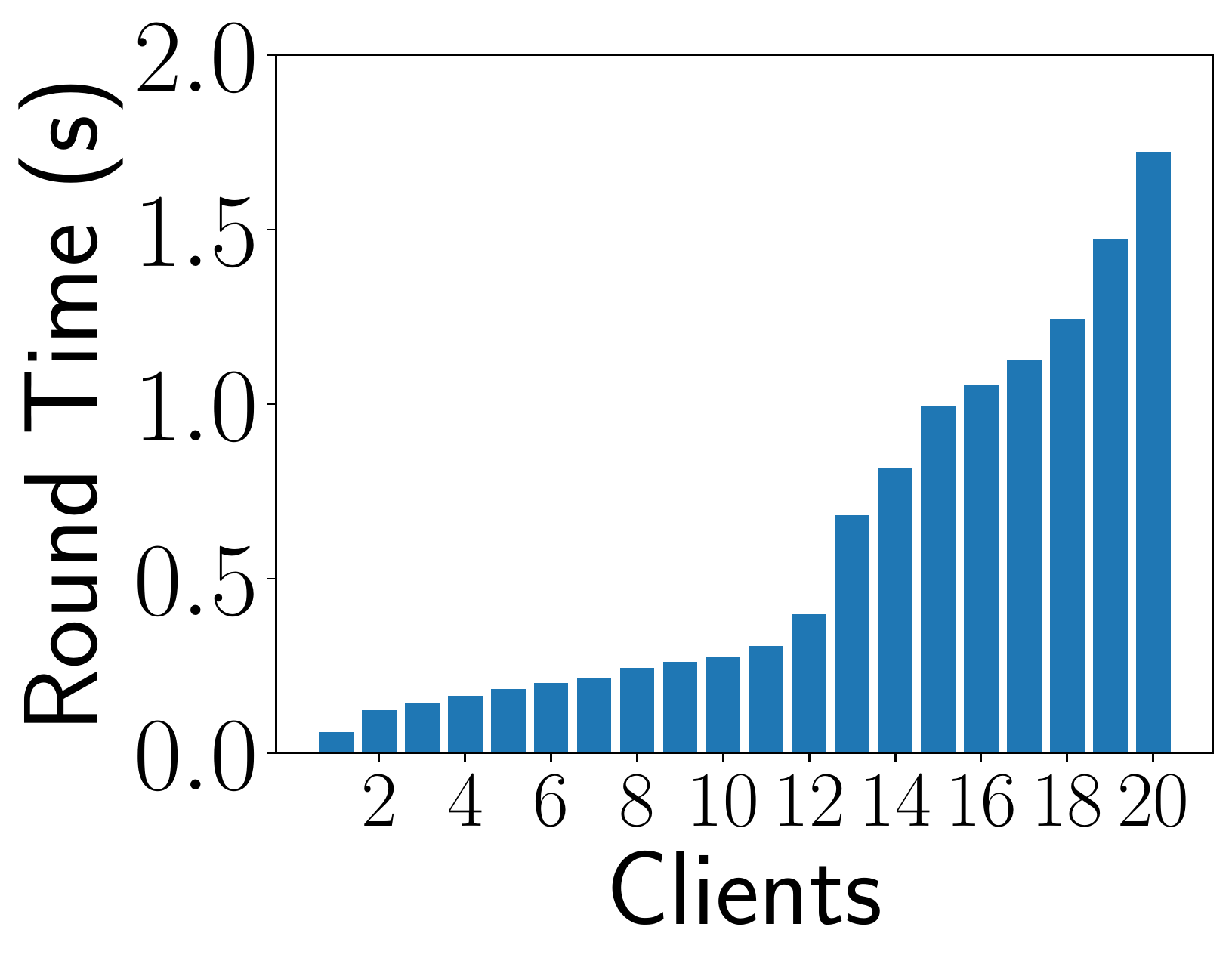}}
  \caption{Impact of heterogeneity simulation on training time of sampled 20 clients in one round of training using FEMNIST dataset: (a) unbalanced data simulated by $Dir(0.5)$, (b) system heterogeneity, and (c) combination effect of (a) and (b). All simulations cause training time variances.}
  \label{fig:heterogenous-impact-femnist}
\end{figure}

\begin{figure}[h]
  \centering
  \subfigure[]{\label{fig:hetero-impact-shakespeare-a}\includegraphics[width=25mm]{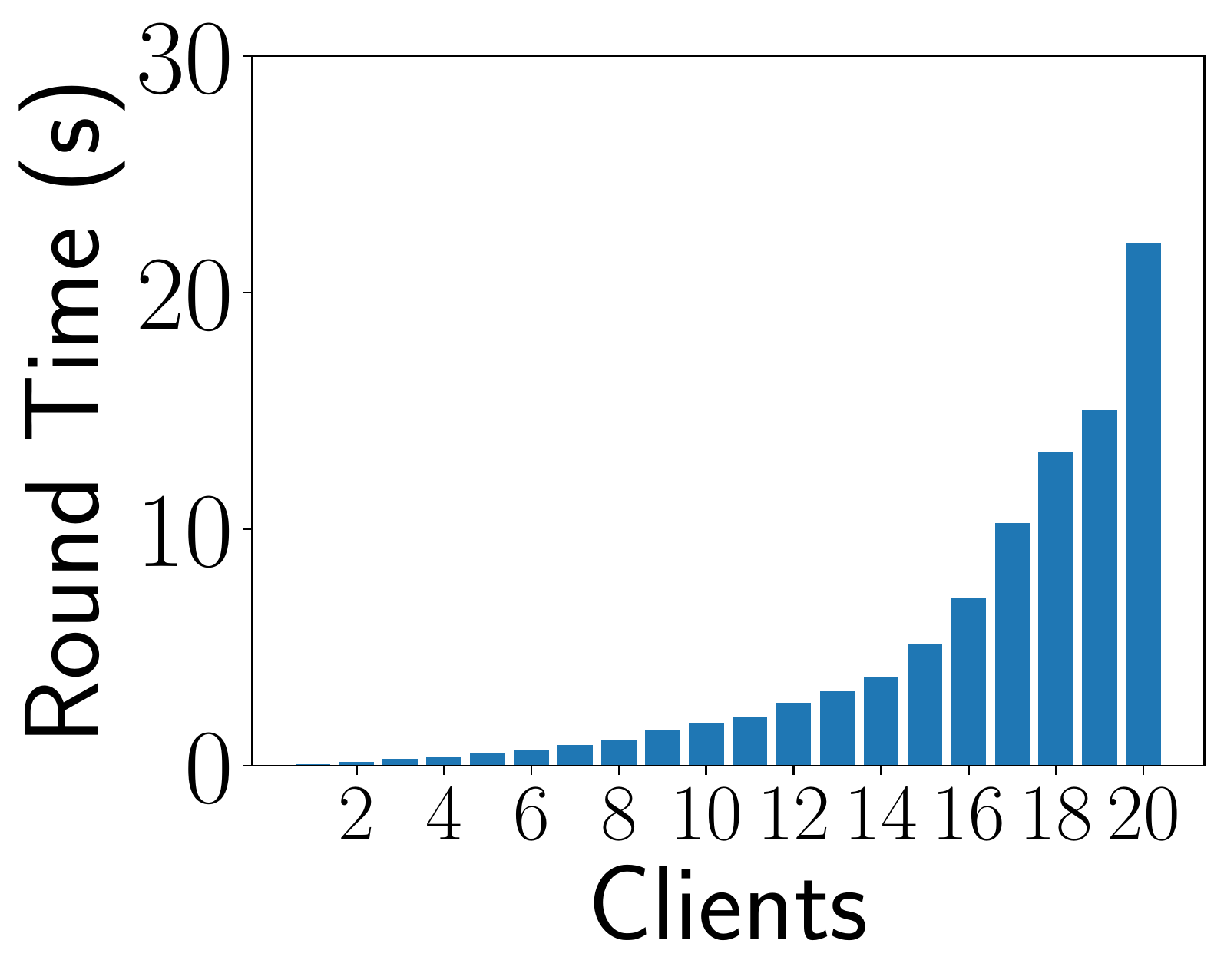}}
  \subfigure[]{\label{fig:hetero-impact-shakespeare-b}\includegraphics[width=25mm]{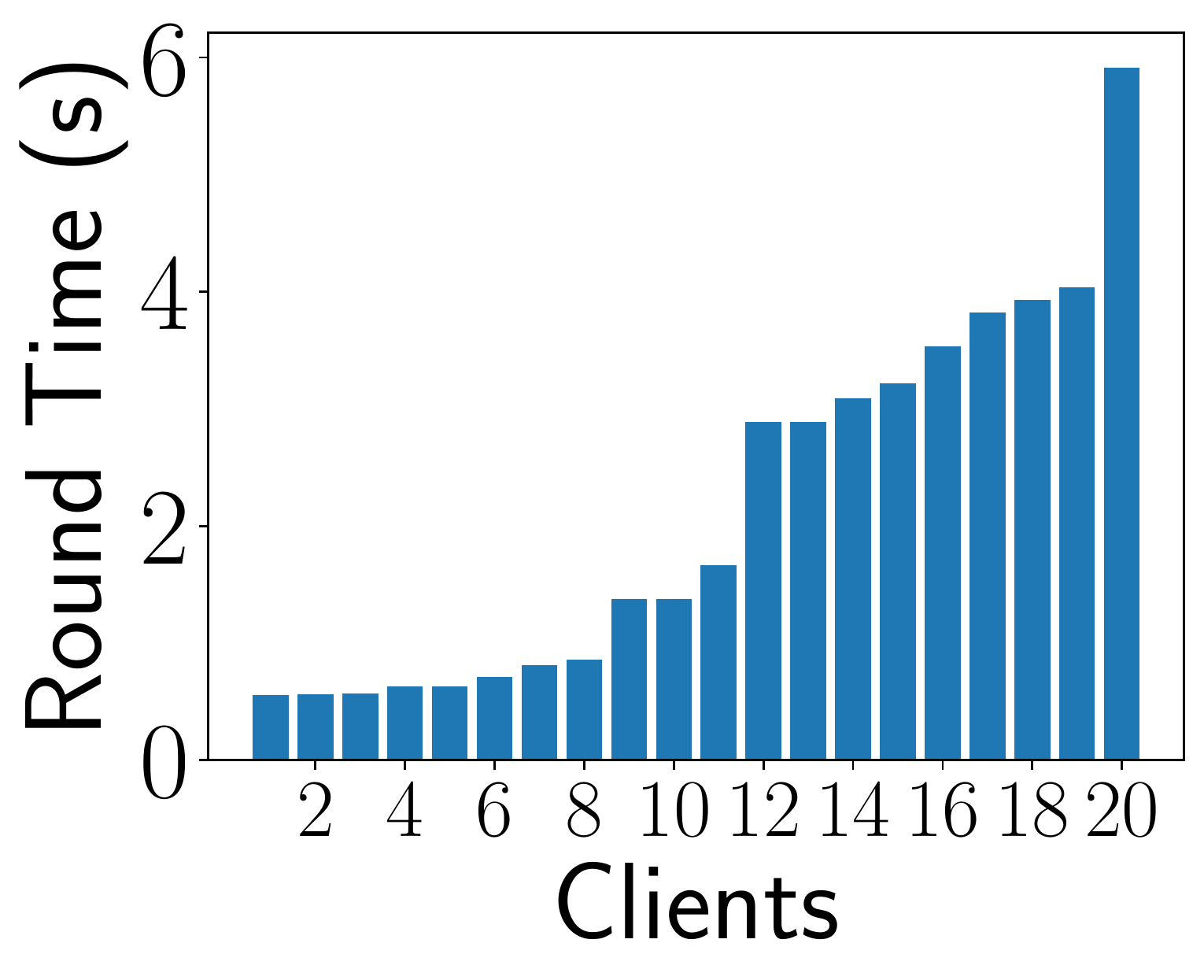}}
  \subfigure[]{\label{fig:hetero-impact-shakespeare-c}\includegraphics[width=25mm]{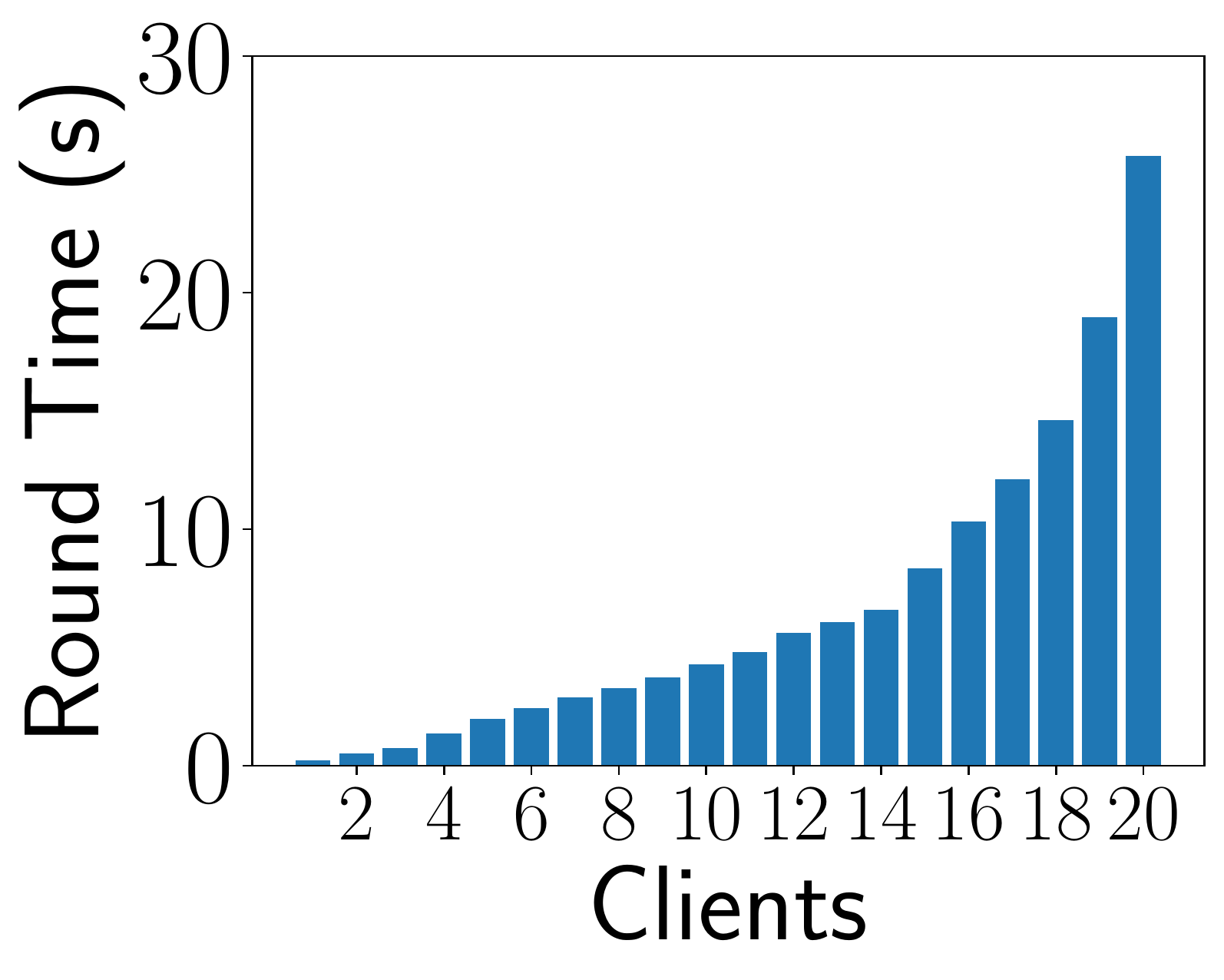}}
  \caption{Impact of heterogeneity simulation on training time of sampled 20 clients in one round of training using Shakespeare dataset: (a) unbalanced data simulated by $Dir(0.5)$, (b) system heterogeneity, and (c) combination effect of (a) and (b). All simulations cause training time variances.}
  \label{fig:heterogenous-impact-shakespeare}
\end{figure}

For experiments on \textit{Performance of GreedyAda} in Section VIII-D, we run experiments with 20 selected clients each round and train FEMNIST with $R =500$ rounds, and Shakespeare and CIFAR-10 with $R = 100$ rounds. 


For experiments on \textit{FedReID}, we use batch size $B = 32$ , $E = 1$, and learning rate setting same as the original paper.

\subsection{Experiment Results}
\label{sec:appx-experiments-results}

\textbf{Impact of Heterogeneity on Training Time} Fig. \ref{fig:heterogenous-impact-femnist} and \ref{fig:heterogenous-impact-shakespeare} show the impact of heterogeneity simulation on training time of each round, using FEMNIST and Shakespeare datasets. They simulate the training time discrepancy and stragglers with unbalanced data and system heterogeneity. The combination of unbalanced data and system heterogeneity simulation results has the largest training time variances. Depending on the datasets, the value and scale of round time are different.

\begin{figure}[t]
  \centering
  \subfigure[FEMNIST]{\label{fig:heterogenous-accuracy-c}\includegraphics[height=25mm]{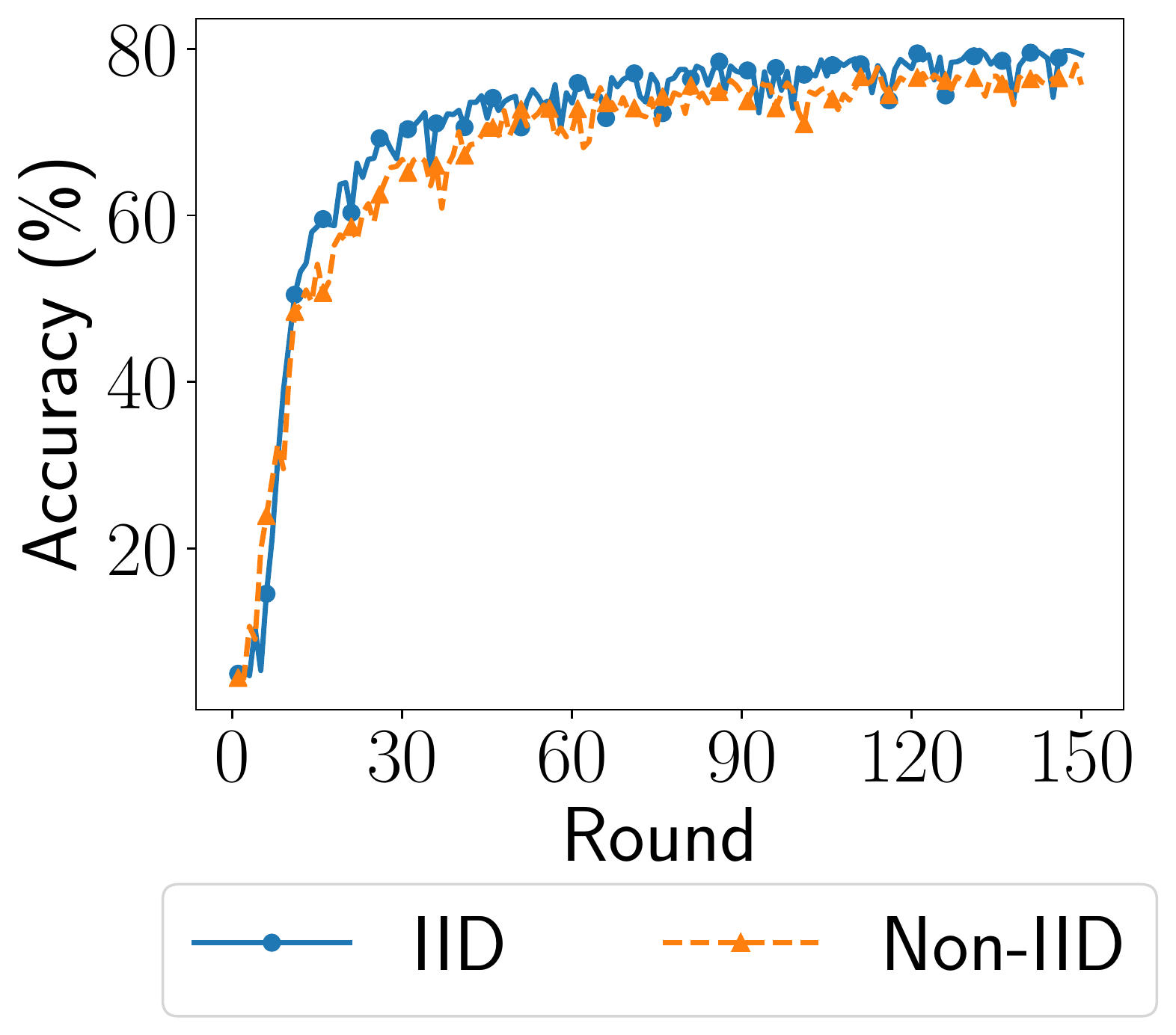}}
  \subfigure[Shakespeare]{\label{fig:heterogenous-accuracy-b}\includegraphics[height=25mm]{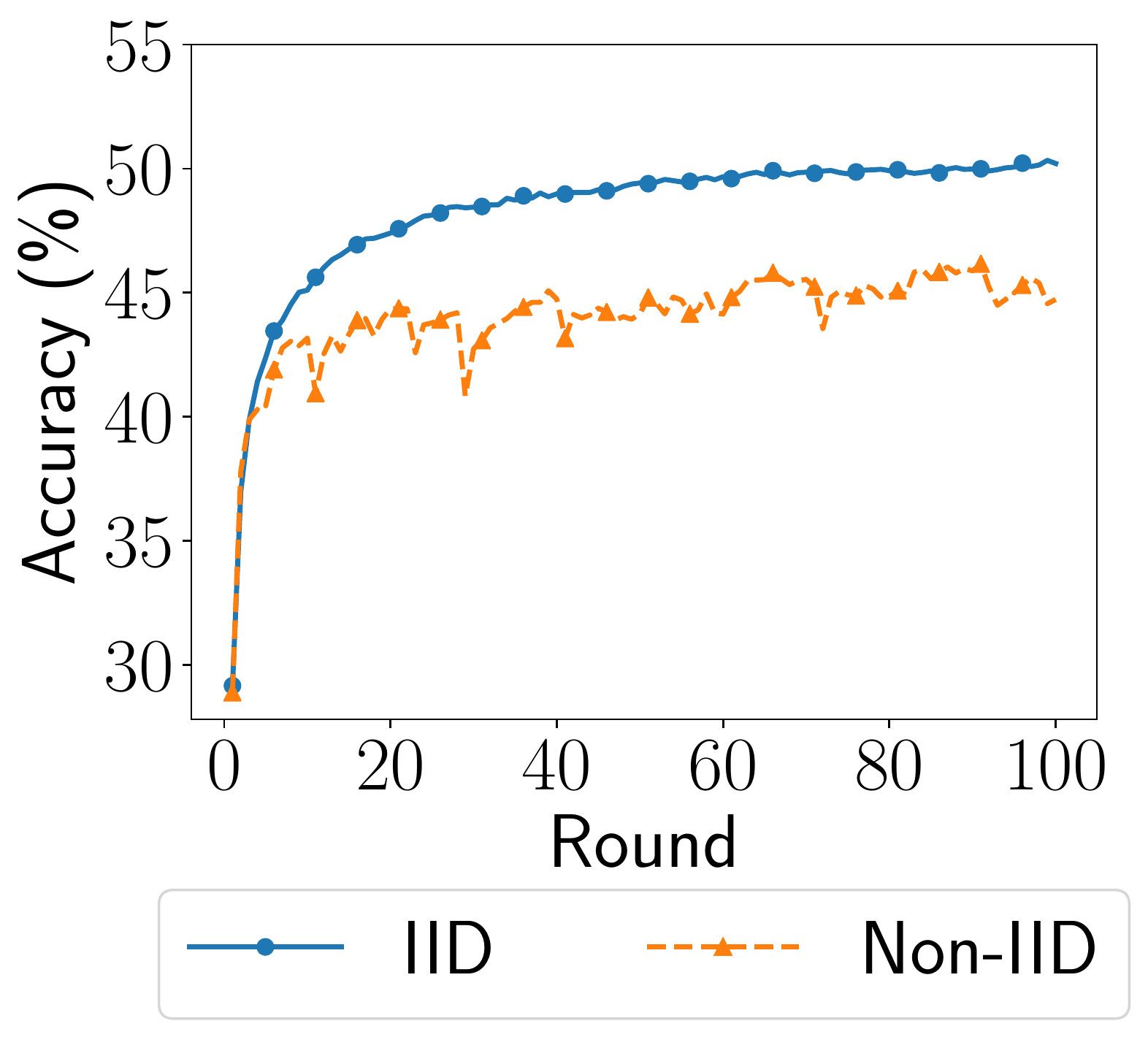}}
  \subfigure[CIFAR-10]{\label{fig:heterogenous-accuracy-a}\includegraphics[height=26mm]{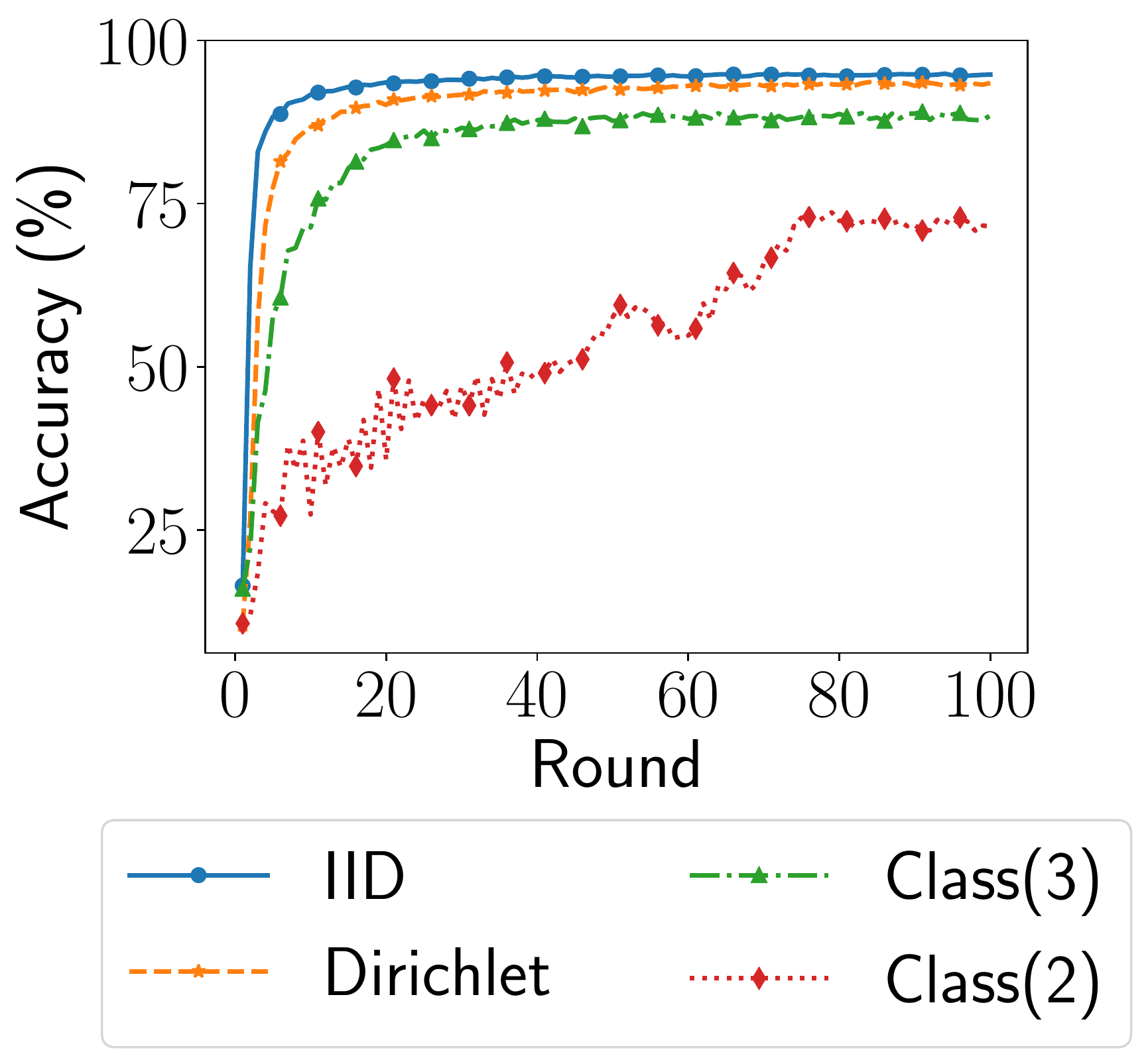}}
  \caption{Performance comparison of IID and Non-IID simulated using EasyFL with selected clients per round $C = 10$. Different non-IID data partition methods lead to different degrees of performance degradation. FEMNIST and Shakespeare use the realistic non-IID partition. CIFAR-10 partitions non-IID by Dirichlet process (dir) and class with $n$ classes per client (class($n$)).}
  \label{fig:heterogenous-accuracy}
\end{figure}

\textbf{Impact of Statistical Heterogeneity} Fig. \ref{fig:heterogenous-accuracy} shows the performance comparison of IID and non-IID data simulated using EasyFL with three datasets: FEMNIST, Shakespeare, and CIFAR-10. These figures further illustrate the accuracy gap between IID and non-IID data.

\section{Recent Publications of Federated Learning}
\label{sec:appx-publications}

We surveyed 33 papers from recent publications of FL from both the machine learning and system community. Table \ref{tab:publications} shows that 10 out of 33 ($\sim$30\%) publications propose new algorithms with changes in only one stage of the training flow, and the majority ($\sim$57\%) change only two stages. Training flow abstraction (Section \ref{sec:workflow-abstraction}) allows researchers to focus on the problems, without re-implementing the whole FL process.

\begingroup
\setlength{\tabcolsep}{0.3em}
\begin{table*}[t]
\caption{Changes in stages of training flow abstraction in recent publications from both machine learning community and system community. Around 30\% of them change only one stage in federated learning and majority change (57\%) only two stages in the FedAvg algorithm.}
\label{tab:publications}
\begin{center}
\begin{tabular}{p{1.7cm}p{6.5cm}cccccccc}
\hline
Conference &
\multicolumn{1}{l}{\multirow{2}{*}{Paper Title}} & 
    \multicolumn{3}{c}{Server} & &
    \multicolumn{3}{c}{Client} \\ \cline{3-5} \cline{7-9}
/ Journal & & Selection & Compression & Aggregation & & Train & Compression & Encryption \\ 
\hline
IoT 2021 & FedMCCS: Multicriteria Client Selection Model for Optimal IoT Federated Learning \cite{abdulrahman2020fedmccs} & $\surd$ & & & & & & \\ \hline
INFOCOM 2020 & Optimizing Federated Learning on Non-IID Data with Reinforcement Learning \cite{wang2020optimizing} & $\surd$ & & & & & & \\ \hline
OSDI 2021 & Oort: Informed Participant Selection for Scalable Federated Learning \cite{lai2020oort} & $\surd$ & & & & & & \\ \hline
HPDC 2020 & TiFL: A Tier-based Federated Learning System \cite{tifl} & $\surd$ & & & & & & \\ \hline
KDD 2020 & FedFast: Going Beyond Average for Faster Training of Federated Recommender Systems \cite{muhammad2020fedfast} & $\surd$ & & $\surd$ & & & & \\ \hline
TNNLS 2019 & Robust and Communication-Efficient Federated Learning From Non-i.i.d. Data \cite{sattler2019robust} & & $\surd$ & & & & $\surd$ & \\ \hline
NIPS 2020 & Ensemble Distillation for Robust Model Fusion in Federated Learning \cite{lin2020ensemble} & & & $\surd$ & & & & \\ \hline
ICDCS 2019 & CMFL: Mitigating Communication Overhead for Federated Learning \cite{luping2019cmfl} & & & & & & $\surd$ & \\ \hline
ICML 2020 & FetchSGD: Communication-Efficient Federated Learning with Sketching \cite{rothchild2020fetchsgd} & & & $\surd$ & & & $\surd$ & \\ \hline
ICML 2020 & SCAFFOLD: Stochastic Controlled Averaging for Federated Learning \cite{karimireddy2020scaffold} & & & $\surd$ & & & $\surd$ & \\ \hline
TPDS 2020 & FedSCR: Structure-Based Communication Reduction for Federated Learning \cite{wu2020fedscr} & & & $\surd$ & & & $\surd$ & \\ \hline
HotEdge 2018 & eSGD: Communication Efficient Distributed Deep Learning on the Edge \cite{tao2018esgd} & & & & & & $\surd$ & \\ \hline
ICML 2020 & Adaptive Federated Optimization \cite{reddi2020adaptive} & & & & & $\surd$ & & \\ \hline
CVPR 2021 & Privacy-preserving Collaborative Learning with Automatic Transformation Search \cite{gao2020privacy} & & & & & $\surd$ & & \\ \hline
MLSys 2020 & Federated Optimization in Heterogeneous Networks \cite{fedprox} & & & & & $\surd$ & & \\ \hline
ICLR 2020 & Federated Learning with Matched Averaging \cite{fedma} & & & $\surd$ & & $\surd$ & & \\ \hline
ACMMM 2020 & Performance Optimization for Federated Person Re-identification via Benchmark Analysis \cite{zhuang2020fedreid} & & & $\surd$ & & $\surd$ & & \\ \hline
NIPS 2020 & Distributionally Robust Federated Averaging \cite{deng2021distributionally} & & & $\surd$ & & $\surd$ & & \\ \hline
NIPS 2020 & Group Knowledge Transfer: Federated Learning of Large CNNs at the Edge \cite{he2020group} & & & $\surd$ & & $\surd$ & & \\ \hline
NIPS 2020 & Personalized Federated Learning with Moreau Envelopes \cite{dinh2020personalized} & & & $\surd$ & & $\surd$ & & \\ \hline
ICLR 2020 & Fair Resource Allocation in Federated Learning \cite{li2019fair} & & & $\surd$ & & $\surd$ & & \\ \hline
ICML 2020 & Federated Learning with Only Positive Labels \cite{felix2020federated} & & & $\surd$ & & $\surd$ & & \\ \hline
AAAI 2021 & Addressing Class Imbalance in Federated Learning \cite{wang2021addressing} & & & $\surd$ & & $\surd$ & & \\ \hline
AAAI 2021 & Federated Block Coordinate Descent Scheme for Learning Global and Personalized Models \cite{wu2020federated} & & & $\surd$ & & $\surd$ & & \\ \hline
IoT 2020 & Toward Communication-Efficient Federated Learning in the Internet of Things With Edge Computing \cite{sun2020toward} & & & $\surd$ & & $\surd$ & $\surd$ & \\ \hline
ICML 2020 & Acceleration for Compressed Gradient Descent in Distributed and Federated Optimization \cite{li2020acceleration} & & & $\surd$ & & $\surd$ & $\surd$ & \\ \hline
INFOCOMM 2018 & When Edge Meets Learning: Adaptive Control for Resource-Constrained Distributed Machine Learning \cite{wang2018edge} & & & $\surd$ & & $\surd$ & & \\ \hline
ATC 2020 & BatchCrypt: Efficient Homomorphic Encryption for Cross-Silo Federated Learning \cite{zhang2020batchcrypt} & & & $\surd$ & & & & $\surd$ \\ \hline
AAAI 2021 & FLAME: Differentially Private Federated Learning in the Shuffle Model \cite{liu2020flame} & & & $\surd$ & & & & $\surd$ \\ \hline
TIFS 2020 & Federated Learning with Differential Privacy: Algorithms and Performance Analysis \cite{wei2020federated} & & & $\surd$ & & & & $\surd$ \\ \hline
GLOBECOM 2020 & Towards Efficient Secure Aggregation for Model Update in Federated Learning \cite{wu2020towards} & & & $\surd$ & & & & $\surd$ \\ \hline
MobiCom 2020 & Billion-Scale Federated Learning on Mobile Clients: A Submodel Design with Tunable Privacy \cite{niu2020billion} & & & $\surd$ & & $\surd$ & & $\surd$ \\ \hline
IoT 2020 & Privacy-Preserving Federated Learning in Fog Computing \cite{zhou2020privacy} & & & $\surd$ & & $\surd$ & & $\surd$ \\ \hline
\end{tabular}
\end{center}
\end{table*}
\endgroup



\ifCLASSOPTIONcaptionsoff
  \newpage
\fi

\bibliographystyle{IEEEtran}
\bibliography{IEEEabrv,references}

\begin{thebibliography}{10}
\providecommand{\url}[1]{#1}
\csname url@samestyle\endcsname
\providecommand{\newblock}{\relax}
\providecommand{\bibinfo}[2]{#2}
\providecommand{\BIBentrySTDinterwordspacing}{\spaceskip=0pt\relax}
\providecommand{\BIBentryALTinterwordstretchfactor}{4}
\providecommand{\BIBentryALTinterwordspacing}{\spaceskip=\fontdimen2\font plus
\BIBentryALTinterwordstretchfactor\fontdimen3\font minus
  \fontdimen4\font\relax}
\providecommand{\BIBforeignlanguage}[2]{{%
\expandafter\ifx\csname l@#1\endcsname\relax
\typeout{** WARNING: IEEEtran.bst: No hyphenation pattern has been}%
\typeout{** loaded for the language `#1'. Using the pattern for}%
\typeout{** the default language instead.}%
\else
\language=\csname l@#1\endcsname
\fi
#2}}
\providecommand{\BIBdecl}{\relax}
\BIBdecl

\bibitem{leaf}
S.~Caldas, S.~M.~K. Duddu, P.~Wu, T.~Li, J.~Kone{\v{c}}n{\`y}, H.~B. McMahan,
  V.~Smith, and A.~Talwalkar, ``Leaf: A benchmark for federated settings,''
  \emph{arXiv preprint arXiv:1812.01097}, 2018.

\bibitem{pysyft}
T.~Ryffel, A.~Trask, M.~Dahl, B.~Wagner, J.~Mancuso, D.~Rueckert, and
  J.~Passerat-Palmbach, ``A generic framework for privacy preserving deep
  learning,'' \emph{arXiv preprint arXiv:1811.04017}, 2018.

\bibitem{ma2019paddlepaddle}
Y.~Ma, D.~Yu, T.~Wu, and H.~Wang, ``Paddlepaddle: An open-source deep learning
  platform from industrial practice,'' \emph{Frontiers of Data and Domputing},
  vol.~1, no.~1, pp. 105--115, 2019.

\bibitem{tff}
\BIBentryALTinterwordspacing
Tensorflow.org, ``Tensorflow federated,'' 2019. [Online]. Available:
  \url{https://github.com/tensorflow/federated}
\BIBentrySTDinterwordspacing

\bibitem{fate}
\BIBentryALTinterwordspacing
WeBank, ``Federated ai technology enabler (fate),'' 2019. [Online]. Available:
  \url{https://github.com/FederatedAI/FATE}
\BIBentrySTDinterwordspacing

\bibitem{sheller2018brain-tumor2}
M.~J. Sheller, G.~A. Reina, B.~Edwards, J.~Martin, and S.~Bakas,
  ``Multi-institutional deep learning modeling without sharing patient data: A
  feasibility study on brain tumor segmentation,'' in \emph{International
  MICCAI Brainlesion Workshop}.\hskip 1em plus 0.5em minus 0.4em\relax
  Springer, 2018, pp. 92--104.

\bibitem{li2019brain-tumor1}
W.~Li, F.~Milletar{\`\i}, D.~Xu, N.~Rieke, J.~Hancox, W.~Zhu, M.~Baust,
  Y.~Cheng, S.~Ourselin, M.~J. Cardoso \emph{et~al.}, ``Privacy-preserving
  federated brain tumour segmentation,'' in \emph{International Workshop on
  Machine Learning in Medical Imaging}.\hskip 1em plus 0.5em minus 0.4em\relax
  Springer, 2019, pp. 133--141.

\bibitem{chen2020fedhealth}
Y.~Chen, X.~Qin, J.~Wang, C.~Yu, and W.~Gao, ``Fedhealth: A federated transfer
  learning framework for wearable healthcare,'' \emph{IEEE Intelligent
  Systems}, 2020.

\bibitem{hard2018gboard}
A.~Hard, K.~Rao, R.~Mathews, S.~Ramaswamy, F.~Beaufays, S.~Augenstein,
  H.~Eichner, C.~Kiddon, and D.~Ramage, ``Federated learning for mobile
  keyboard prediction,'' \emph{arXiv preprint arXiv:1811.03604}, 2018.

\bibitem{leroy2019kwakeword}
D.~Leroy, A.~Coucke, T.~Lavril, T.~Gisselbrecht, and J.~Dureau, ``Federated
  learning for keyword spotting,'' in \emph{ICASSP 2019-2019 IEEE International
  Conference on Acoustics, Speech and Signal Processing (ICASSP)}.\hskip 1em
  plus 0.5em minus 0.4em\relax IEEE, 2019, pp. 6341--6345.

\bibitem{niu2020recommendation-alibaba}
C.~Niu, F.~Wu, S.~Tang, L.~Hua, R.~Jia, C.~Lv, Z.~Wu, and G.~Chen,
  ``Billion-scale federated learning on mobile clients: A submodel design with
  tunable privacy,'' in \emph{Proceedings of the 26th Annual International
  Conference on Mobile Computing and Networking}, 2020, pp. 1--14.

\bibitem{muhammad2020fedfast}
K.~Muhammad, Q.~Wang, D.~O'Reilly-Morgan, E.~Tragos, B.~Smyth, N.~Hurley,
  J.~Geraci, and A.~Lawlor, ``Fedfast: Going beyond average for faster training
  of federated recommender systems,'' in \emph{Proceedings of the 26th ACM
  SIGKDD International Conference on Knowledge Discovery \& Data Mining}, 2020,
  pp. 1234--1242.

\bibitem{zhuang2020fedreid}
W.~Zhuang, Y.~Wen, X.~Zhang, X.~Gan, D.~Yin, D.~Zhou, S.~Zhang, and S.~Yi,
  ``Performance optimization of federated person re-identification via
  benchmark analysis,'' in \emph{Proceedings of the 28th ACM International
  Conference on Multimedia}, 2020, pp. 955--963.

\bibitem{yang2018gboard}
T.~Yang, G.~Andrew, H.~Eichner, H.~Sun, W.~Li, N.~Kong, D.~Ramage, and
  F.~Beaufays, ``Applied federated learning: Improving google keyboard query
  suggestions,'' \emph{arXiv preprint arXiv:1812.02903}, 2018.

\bibitem{rieke2020nvidia-fl-health}
N.~Rieke, J.~Hancox, W.~Li, F.~Milletari, H.~R. Roth, S.~Albarqouni, S.~Bakas,
  M.~N. Galtier, B.~A. Landman, K.~Maier-Hein \emph{et~al.}, ``The future of
  digital health with federated learning,'' \emph{NPJ digital medicine},
  vol.~3, no.~1, pp. 1--7, 2020.

\bibitem{gartner}
\BIBentryALTinterwordspacing
K.~Panetta. (2020) Gartner top strategic technology trends for 2021. [Online].
  Available:
  \url{https://www.gartner.com/smarterwithgartner/gartner-top-strategic-technology-trends-for-2021/}
\BIBentrySTDinterwordspacing

\bibitem{abadi2016tensorflow}
M.~Abadi, P.~Barham, J.~Chen, Z.~Chen, A.~Davis, J.~Dean, M.~Devin,
  S.~Ghemawat, G.~Irving, M.~Isard \emph{et~al.}, ``Tensorflow: A system for
  large-scale machine learning,'' in \emph{12th $\{$USENIX$\}$ symposium on
  operating systems design and implementation ($\{$OSDI$\}$ 16)}, 2016, pp.
  265--283.

\bibitem{chollet2015keras}
F.~Chollet \emph{et~al.}, ``Keras,'' \url{https://keras.io}, 2015.

\bibitem{kairouz2019fl-advances-open}
P.~Kairouz, H.~B. McMahan, B.~Avent, A.~Bellet, M.~Bennis, A.~N. Bhagoji,
  K.~Bonawitz, Z.~Charles, G.~Cormode, R.~Cummings \emph{et~al.}, ``Advances
  and open problems in federated learning,'' \emph{arXiv preprint
  arXiv:1912.04977}, 2019.

\bibitem{Li2020FedChallenges}
T.~Li, A.~K. Sahu, A.~Talwalkar, and V.~Smith, ``Federated learning:
  Challenges, methods, and future directions,'' \emph{IEEE Signal Processing
  Magazine}, vol.~37, pp. 50--60, 2020.

\bibitem{fedavg}
B.~McMahan, E.~Moore, D.~Ramage, S.~Hampson, and B.~A. y~Arcas,
  ``Communication-efficient learning of deep networks from decentralized
  data,'' in \emph{Artificial Intelligence and Statistics}.\hskip 1em plus
  0.5em minus 0.4em\relax PMLR, 2017, pp. 1273--1282.

\bibitem{zhao2018non-iid}
Y.~Zhao, M.~Li, L.~Lai, N.~Suda, D.~Civin, and V.~Chandra, ``Federated learning
  with non-iid data,'' \emph{arXiv preprint arXiv:1806.00582}, 2018.

\bibitem{fedprox}
T.~Li, A.~K. Sahu, M.~Zaheer, M.~Sanjabi, A.~Talwalkar, and V.~Smith,
  ``Federated optimization in heterogeneous networks,'' in \emph{Proceedings of
  Machine Learning and Systems 2020}, 2020, pp. 429--450.

\bibitem{tifl}
Z.~Chai, A.~Ali, S.~Zawad, S.~Truex, A.~Anwar, N.~Baracaldo, Y.~Zhou,
  H.~Ludwig, F.~Yan, and Y.~Cheng, ``Tifl: A tier-based federated learning
  system,'' in \emph{Proceedings of the 29th International Symposium on
  High-Performance Parallel and Distributed Computing}, 2020, pp. 125--136.

\bibitem{flash}
C.~Yang, Q.~Wang, M.~Xu, S.~Wang, K.~Bian, and X.~Liu, ``Heterogeneity-aware
  federated learning,'' \emph{arXiv preprint arXiv:2006.06983}, 2020.

\bibitem{large-scale-dnn-NIPS2012}
J.~Dean, G.~Corrado, R.~Monga, K.~Chen, M.~Devin, M.~Mao, M.~Ranzato,
  A.~Senior, P.~Tucker, K.~Yang \emph{et~al.}, ``Large scale distributed deep
  networks,'' in \emph{Advances in neural information processing systems},
  2012, pp. 1223--1231.

\bibitem{ml-debt}
D.~Sculley, G.~Holt, D.~Golovin, E.~Davydov, T.~Phillips, D.~Ebner,
  V.~Chaudhary, and M.~Young, ``Machine learning: The high interest credit card
  of technical debt,'' in \emph{SE4ML: Software Engineering for Machine
  Learning (NIPS 2014 Workshop)}, 2014.

\bibitem{burkov2020mle-book}
A.~Burkov, \emph{Machine learning engineering}.\hskip 1em plus 0.5em minus
  0.4em\relax True Positive Incorporated, 2020.

\bibitem{zhang2020mlmodelci}
H.~Zhang, Y.~Li, Y.~Huang, Y.~Wen, J.~Yin, and K.~Guan, ``Mlmodelci: An
  automatic cloud platform for efficient mlaas,'' in \emph{Proceedings of the
  28th ACM International Conference on Multimedia}, 2020, pp. 4453--4456.

\bibitem{karlavs2020mlops-kdd}
B.~Karla{\v{s}}, M.~Interlandi, C.~Renggli, W.~Wu, C.~Zhang, D.~Mukunthu
  Iyappan~Babu, J.~Edwards, C.~Lauren, A.~Xu, and M.~Weimer, ``Building
  continuous integration services for machine learning,'' in \emph{Proceedings
  of the 26th ACM SIGKDD International Conference on Knowledge Discovery \&
  Data Mining}, 2020, pp. 2407--2415.

\bibitem{Bonawitz2019FL-sys-scale}
\BIBentryALTinterwordspacing
K.~Bonawitz, H.~Eichner, W.~Grieskamp, D.~Huba, A.~Ingerman, V.~Ivanov,
  C.~Kiddon, J.~Kone\v{c}n\'{y}, S.~Mazzocchi, B.~McMahan, T.~Van~Overveldt,
  D.~Petrou, D.~Ramage, and J.~Roselander, ``Towards federated learning at
  scale: System design,'' in \emph{Proceedings of Machine Learning and
  Systems}, A.~Talwalkar, V.~Smith, and M.~Zaharia, Eds., 2019, vol.~1, pp.
  374--388. [Online]. Available:
  \url{https://proceedings.mlsys.org/paper/2019/file/bd686fd640be98efaae0091fa301e613-Paper.pdf}
\BIBentrySTDinterwordspacing

\bibitem{cohen2017emnist}
G.~Cohen, S.~Afshar, J.~Tapson, and A.~Van~Schaik, ``Emnist: Extending mnist to
  handwritten letters,'' in \emph{2017 International Joint Conference on Neural
  Networks (IJCNN)}.\hskip 1em plus 0.5em minus 0.4em\relax IEEE, 2017, pp.
  2921--2926.

\bibitem{shakespeare2007shakespeare}
W.~Shakespeare, \emph{The complete works of William Shakespeare}.\hskip 1em
  plus 0.5em minus 0.4em\relax Wordsworth Editions, 2007.

\bibitem{cifar2009}
A.~Krizhevsky, G.~Hinton \emph{et~al.}, ``Learning multiple layers of features
  from tiny images,'' 2009.

\bibitem{fedma}
\BIBentryALTinterwordspacing
H.~Wang, M.~Yurochkin, Y.~Sun, D.~Papailiopoulos, and Y.~Khazaeni, ``Federated
  learning with matched averaging,'' in \emph{International Conference on
  Learning Representations}, 2020. [Online]. Available:
  \url{https://openreview.net/forum?id=BkluqlSFDS}
\BIBentrySTDinterwordspacing

\bibitem{imagenet_cvpr09}
J.~Deng, W.~Dong, R.~Socher, L.-J. Li, K.~Li, and L.~Fei-Fei, ``{ImageNet: A
  Large-Scale Hierarchical Image Database},'' in \emph{CVPR09}, 2009.

\bibitem{ignatov2018aibenchmark}
A.~Ignatov, R.~Timofte, W.~Chou, K.~Wang, M.~Wu, T.~Hartley, and L.~Van~Gool,
  ``Ai benchmark: Running deep neural networks on android smartphones,'' in
  \emph{Proceedings of the European conference on computer vision (ECCV)},
  2018, pp. 0--0.

\bibitem{sattler2019stc}
F.~Sattler, S.~Wiedemann, K.-R. M{\"u}ller, and W.~Samek, ``Robust and
  communication-efficient federated learning from non-iid data,'' \emph{IEEE
  transactions on neural networks and learning systems}, 2019.

\bibitem{lai2020oort}
F.~Lai, X.~Zhu, H.~V. Madhyastha, and M.~Chowdhury, ``Oort: Informed
  participant selection for scalable federated learning,'' \emph{arXiv preprint
  arXiv:2010.06081}, 2020.

\bibitem{graham1969bounds}
R.~L. Graham, ``Bounds on multiprocessing timing anomalies,'' \emph{SIAM
  journal on Applied Mathematics}, vol.~17, no.~2, pp. 416--429, 1969.

\bibitem{johnson1974worst}
D.~S. Johnson, A.~Demers, J.~D. Ullman, M.~R. Garey, and R.~L. Graham,
  ``Worst-case performance bounds for simple one-dimensional packing
  algorithms,'' \emph{SIAM Journal on computing}, vol.~3, no.~4, pp. 299--325,
  1974.

\bibitem{docker}
\BIBentryALTinterwordspacing
Docker, ``Docker,'' 2013. [Online]. Available: \url{https://www.docker.com/}
\BIBentrySTDinterwordspacing

\bibitem{k8s}
\BIBentryALTinterwordspacing
Kubernetes.io, ``Kubernetes,'' 2014. [Online]. Available:
  \url{https://kubernetes.io/}
\BIBentrySTDinterwordspacing

\bibitem{paszke2017pytorch}
A.~Paszke, S.~Gross, S.~Chintala, G.~Chanan, E.~Yang, Z.~DeVito, Z.~Lin,
  A.~Desmaison, L.~Antiga, and A.~Lerer, ``Automatic differentiation in
  pytorch,'' 2017.

\bibitem{paszke2019pytorch}
A.~Paszke, S.~Gross, F.~Massa, A.~Lerer, J.~Bradbury, G.~Chanan, T.~Killeen,
  Z.~Lin, N.~Gimelshein, L.~Antiga \emph{et~al.}, ``Pytorch: An imperative
  style, high-performance deep learning library,'' \emph{Advances in neural
  information processing systems}, vol.~32, pp. 8026--8037, 2019.

\bibitem{abdulrahman2020fedmccs}
S.~AbdulRahman, H.~Tout, A.~Mourad, and C.~Talhi, ``Fedmccs: multicriteria
  client selection model for optimal iot federated learning,'' \emph{IEEE
  Internet of Things Journal}, vol.~8, no.~6, pp. 4723--4735, 2020.

\bibitem{wang2020optimizing}
H.~Wang, Z.~Kaplan, D.~Niu, and B.~Li, ``Optimizing federated learning on
  non-iid data with reinforcement learning,'' in \emph{IEEE INFOCOM 2020-IEEE
  Conference on Computer Communications}.\hskip 1em plus 0.5em minus
  0.4em\relax IEEE, 2020, pp. 1698--1707.

\bibitem{sattler2019robust}
F.~Sattler, S.~Wiedemann, K.-R. M{\"u}ller, and W.~Samek, ``Robust and
  communication-efficient federated learning from non-iid data,'' \emph{IEEE
  transactions on neural networks and learning systems}, vol.~31, no.~9, pp.
  3400--3413, 2019.

\bibitem{lin2020ensemble}
T.~Lin, L.~Kong, S.~U. Stich, and M.~Jaggi, ``Ensemble distillation for robust
  model fusion in federated learning,'' \emph{arXiv preprint arXiv:2006.07242},
  2020.

\bibitem{luping2019cmfl}
W.~Luping, W.~Wei, and L.~Bo, ``Cmfl: Mitigating communication overhead for
  federated learning,'' in \emph{2019 IEEE 39th International Conference on
  Distributed Computing Systems (ICDCS)}.\hskip 1em plus 0.5em minus
  0.4em\relax IEEE, 2019, pp. 954--964.

\bibitem{rothchild2020fetchsgd}
D.~Rothchild, A.~Panda, E.~Ullah, N.~Ivkin, I.~Stoica, V.~Braverman,
  J.~Gonzalez, and R.~Arora, ``Fetchsgd: Communication-efficient federated
  learning with sketching,'' in \emph{International Conference on Machine
  Learning}.\hskip 1em plus 0.5em minus 0.4em\relax PMLR, 2020, pp. 8253--8265.

\bibitem{karimireddy2020scaffold}
S.~P. Karimireddy, S.~Kale, M.~Mohri, S.~Reddi, S.~Stich, and A.~T. Suresh,
  ``Scaffold: Stochastic controlled averaging for federated learning,'' in
  \emph{International Conference on Machine Learning}.\hskip 1em plus 0.5em
  minus 0.4em\relax PMLR, 2020, pp. 5132--5143.

\bibitem{wu2020fedscr}
X.~Wu, X.~Yao, and C.-L. Wang, ``Fedscr: Structure-based communication
  reduction for federated learning,'' \emph{IEEE Transactions on Parallel and
  Distributed Systems}, 2020.

\bibitem{tao2018esgd}
Z.~Tao and Q.~Li, ``esgd: Communication efficient distributed deep learning on
  the edge,'' in \emph{$\{$USENIX$\}$ Workshop on Hot Topics in Edge Computing
  (HotEdge 18)}, 2018.

\bibitem{reddi2020adaptive}
S.~Reddi, Z.~Charles, M.~Zaheer, Z.~Garrett, K.~Rush, J.~Kone{\v{c}}n{\`y},
  S.~Kumar, and H.~B. McMahan, ``Adaptive federated optimization,'' \emph{arXiv
  preprint arXiv:2003.00295}, 2020.

\bibitem{gao2020privacy}
W.~Gao, S.~Guo, T.~Zhang, H.~Qiu, Y.~Wen, and Y.~Liu, ``Privacy-preserving
  collaborative learning with automatic transformation search,'' \emph{arXiv
  preprint arXiv:2011.12505}, 2020.

\bibitem{deng2021distributionally}
Y.~Deng, M.~M. Kamani, and M.~Mahdavi, ``Distributionally robust federated
  averaging,'' \emph{arXiv preprint arXiv:2102.12660}, 2021.

\bibitem{he2020group}
C.~He, M.~Annavaram, and S.~Avestimehr, ``Group knowledge transfer: Federated
  learning of large cnns at the edge,'' \emph{Advances in Neural Information
  Processing Systems}, vol.~33, 2020.

\bibitem{dinh2020personalized}
C.~T. Dinh, N.~H. Tran, and T.~D. Nguyen, ``Personalized federated learning
  with moreau envelopes,'' \emph{arXiv preprint arXiv:2006.08848}, 2020.

\bibitem{li2019fair}
T.~Li, M.~Sanjabi, A.~Beirami, and V.~Smith, ``Fair resource allocation in
  federated learning,'' \emph{arXiv preprint arXiv:1905.10497}, 2019.

\bibitem{felix2020federated}
X.~Y. Felix, A.~S. Rawat, A.~K. Menon, and S.~Kumar, ``Federated learning with
  only positive labels.'' \emph{arXiv preprint arXiv:2004.10342}, 2020.

\bibitem{wang2021addressing}
L.~Wang, S.~Xu, X.~Wang, and Q.~Zhu, ``Addressing class imbalance in federated
  learning,'' \emph{arXiv preprint arXiv:2008.06217}, 2021.

\bibitem{wu2020federated}
R.~Wu, A.~Scaglione, H.-T. Wai, N.~Karakoc, K.~Hreinsson, and W.-K. Ma,
  ``Federated block coordinate descent scheme for learning global and
  personalized models,'' \emph{arXiv preprint arXiv:2012.13900}, 2020.

\bibitem{sun2020toward}
H.~Sun, S.~Li, F.~R. Yu, Q.~Qi, J.~Wang, and J.~Liao, ``Toward
  communication-efficient federated learning in the internet of things with
  edge computing,'' \emph{IEEE Internet of Things Journal}, vol.~7, no.~11, pp.
  11\,053--11\,067, 2020.

\bibitem{li2020acceleration}
Z.~Li, D.~Kovalev, X.~Qian, and P.~Richt{\'a}rik, ``Acceleration for compressed
  gradient descent in distributed and federated optimization,'' \emph{arXiv
  preprint arXiv:2002.11364}, 2020.

\bibitem{wang2018edge}
S.~Wang, T.~Tuor, T.~Salonidis, K.~K. Leung, C.~Makaya, T.~He, and K.~Chan,
  ``When edge meets learning: Adaptive control for resource-constrained
  distributed machine learning,'' in \emph{IEEE INFOCOM 2018-IEEE Conference on
  Computer Communications}.\hskip 1em plus 0.5em minus 0.4em\relax IEEE, 2018,
  pp. 63--71.

\bibitem{zhang2020batchcrypt}
C.~Zhang, S.~Li, J.~Xia, W.~Wang, F.~Yan, and Y.~Liu, ``Batchcrypt: Efficient
  homomorphic encryption for cross-silo federated learning,'' in \emph{2020
  $\{$USENIX$\}$ Annual Technical Conference ($\{$USENIX$\}$$\{$ATC$\}$ 20)},
  2020, pp. 493--506.

\bibitem{liu2020flame}
R.~Liu, Y.~Cao, H.~Chen, R.~Guo, and M.~Yoshikawa, ``Flame: Differentially
  private federated learning in the shuffle model,'' \emph{arXiv preprint
  arXiv:2009.08063}, 2020.

\bibitem{wei2020federated}
K.~Wei, J.~Li, M.~Ding, C.~Ma, H.~H. Yang, F.~Farokhi, S.~Jin, T.~Q. Quek, and
  H.~V. Poor, ``Federated learning with differential privacy: Algorithms and
  performance analysis,'' \emph{IEEE Transactions on Information Forensics and
  Security}, vol.~15, pp. 3454--3469, 2020.

\bibitem{wu2020towards}
D.~Wu, M.~Pan, Z.~Xu, Y.~Zhang, and Z.~Han, ``Towards efficient secure
  aggregation for model update in federated learning,'' in \emph{GLOBECOM
  2020-2020 IEEE Global Communications Conference}.\hskip 1em plus 0.5em minus
  0.4em\relax IEEE, 2020, pp. 1--6.

\bibitem{niu2020billion}
C.~Niu, F.~Wu, S.~Tang, L.~Hua, R.~Jia, C.~Lv, Z.~Wu, and G.~Chen,
  ``Billion-scale federated learning on mobile clients: A submodel design with
  tunable privacy,'' in \emph{Proceedings of the 26th Annual International
  Conference on Mobile Computing and Networking}, 2020, pp. 1--14.

\bibitem{zhou2020privacy}
C.~Zhou, A.~Fu, S.~Yu, W.~Yang, H.~Wang, and Y.~Zhang, ``Privacy-preserving
  federated learning in fog computing,'' \emph{IEEE Internet of Things
  Journal}, vol.~7, no.~11, pp. 10\,782--10\,793, 2020.

\end{thebibliography}
\end{document}